\documentclass[aps,pra,twocolumn,notitlepage,superscriptaddress,showpacs,10pt,longbibliography]{revtex4-2} 
\usepackage{amsmath}
\usepackage{amssymb}
\usepackage{txfonts}
\usepackage{pxfonts}
\usepackage{graphicx,bm,units,yfonts}
\usepackage[table]{xcolor}
\usepackage{hyperref}
\usepackage{color}
\usepackage[utf8]{inputenc}

\usepackage{amstext}
\usepackage{physics}

\makeatletter
\let\openbox\@undefined
\makeatother
\usepackage{amsthm}
\usepackage{amssymb}

\usepackage{appendix}
\makeatletter
\theoremstyle{plain}
\newtheorem{thm}{\protect\theoremname}[section]
\theoremstyle{remark}

\usepackage{xr}

\makeatother

\providecommand{\remarkname}{Remark}
\providecommand{\theoremname}{Theorem}

\begin{document}
\title{Backscattering-free edge states below all bands in two-dimensional auxetic media}

\author{Wenting Cheng}
\email{cwenting@umich.edu}
\thanks{These three authors contributed equally}
\affiliation{Department of Physics, University of Michigan, Ann Arbor, MI 48109, USA}

\author{Kai Qian}
\email{k3qian@ucsd.edu}
\thanks{These three authors contributed equally}
\affiliation{Department of Mechanical and Aerospace Engineering, University of California San Diego, La Jolla, CA 92093, USA}

\author{Nan Cheng}
\email{nancheng@umich.edu}
\thanks{These three authors contributed equally}
\affiliation{Department of Physics, University of Michigan, Ann Arbor, MI 48109, USA}

\author{Nicholas Boechler}
\email{nboechler@ucsd.edu}
\affiliation{Department of Mechanical and Aerospace Engineering, University of California San Diego, La Jolla, CA 92093, USA}
\affiliation{Program in Materials Science and Engineering, University of California San Diego, La Jolla, CA 92093, USA}

\author{Xiaoming Mao}
\email{maox@umich.edu}
\affiliation{Department of Physics, University of Michigan, Ann Arbor, MI 48109, USA}

\author{Kai Sun}
\email{sunkai@umich.edu}
\affiliation{Department of Physics, University of Michigan, Ann Arbor, MI 48109, USA}

\begin{abstract}
Unidirectional and backscattering-free propagation of sound waves is of fundamental interest in physics, and highly sought-after in engineering. Current strategies utilize topologically protected chiral edge modes in bandgaps, or complex mechanisms involving active constituents or nonlinearity. Here we propose a new class of passive, linear, one-way edge states based on spin-momentum locking of Rayleigh waves in two-dimensional media in the limit of vanishing bulk modulus, 
which provides $100\%$ unidirectional and backscattering-free edge propagation at a broad range of frequencies instead of residing in gaps between bulk bands.  We further show that such modes are characterized by a new topological winding number that is analogous to discrete angular momentum eigenvalues in quantum mechanics. These passive and backscattering-free edge waves have the potential to enable a new class of phononic devices in the form of lattices or continua that work in previously inaccessible frequency ranges.
\end{abstract}

\maketitle

\newpage

\newpage

\section{Introduction}

Topologically-protected chiral edge modes have been the focus of extensive studies over the past few decades, where nontrivial topological structures of bulk waves give rise to edge and interface waves exhibiting features such as one-way transport, quantized transport coefficients, and robustness against disorder
~\cite{thouless1982quantized,sheng2005nondissipative, prodan2009robustness, sheng2006quantum, teo2008surface, takahashi2011gapless, zhang2011spontaneous, ezawa2013spin}. 
These intriguing phenomena have been observed in a range of systems from quantum electronic topological materials to classical phononic 
or photonic metamaterials~\cite{haldane2008possible, raghu2008analogs, wang2008reflection, wang2009observation, prodan2009topological, nash2015topological,Wang2015}. 
However, such exotic physics is subject to one constraint: these edge modes (including quantum Hall and  spin, valley, or mirror Hall types) can only arise in frequency gaps between bulk bands. 

Is it possible to break this constraint? 
For quantum Hall edge states, this is impossible, because they are associated with a gauge anomaly, which cannot be achieved below the lowest band (or above the highest)~\cite{thouless1982quantized,Haldane1988}. 
In contrast, one-way transport of topological edge modes of   spin, valley, or mirror Hall types 
come from (pseudo) spin-momentum locking~\cite{thouless1982quantized,sheng2005nondissipative, prodan2009robustness, sheng2006quantum, teo2008surface, takahashi2011gapless, zhang2011spontaneous, ezawa2013spin}. Thus, if the same type of spin-momentum locking can be realized through a different mechanism, it is in principle possible to achieve similar physical properties without the in-gap constraint.

For sound waves, partial spin-momentum locking has long been known~\cite{lipsett1988reexamination,long2018intrinsic, shi2019observation, bliokh2019transverse, long2020symmetry, long2020realization}.
One example is Rayleigh waves \cite{Rayleigh1887} on the surface of a solid, where the surface exhibits elliptical motions (also known as ``ground roll" in seismology \cite{telford1990applied}). 
Interestingly, waves propagating in opposite directions exhibit opposite rolling polarizations. 
The physics  behind this is the breaking of inversion symmetry. 
Rotations and linear motions are characterized by two different types of physical quantities: pseudo-vectors (\textit{e.g.}, angular momentum) and vectors (\textit{e.g.}, linear momentum), which can only couple when inversion symmetry is broken, due to their opposite parity. 
The presence of a physical boundary
serves  this role of breaking symmetry, and thus allows these two different types of motions to be coupled.
Although such locking naturally arises for surface modes, this is generally a partial locking, in contrast to topological edge modes of quantum spin~\cite{susstrunk2015observation, mousavi2015topologically, miniaci2018experimental} or valley~\cite{lu2016valley, lu2017observation, pal2017edge, vila2017observation, zhu2018design, liu2018tunable, wu2018dial, chaunsali2018subwavelength} Hall systems, where the locking is 100\%. 
As with all other elliptically polarized waves, Rayleigh waves contain both left- and right-circularly polarized components, just with different amplitudes and phases.

\begin{figure*}[ht]
\centering
\includegraphics[width=\linewidth]{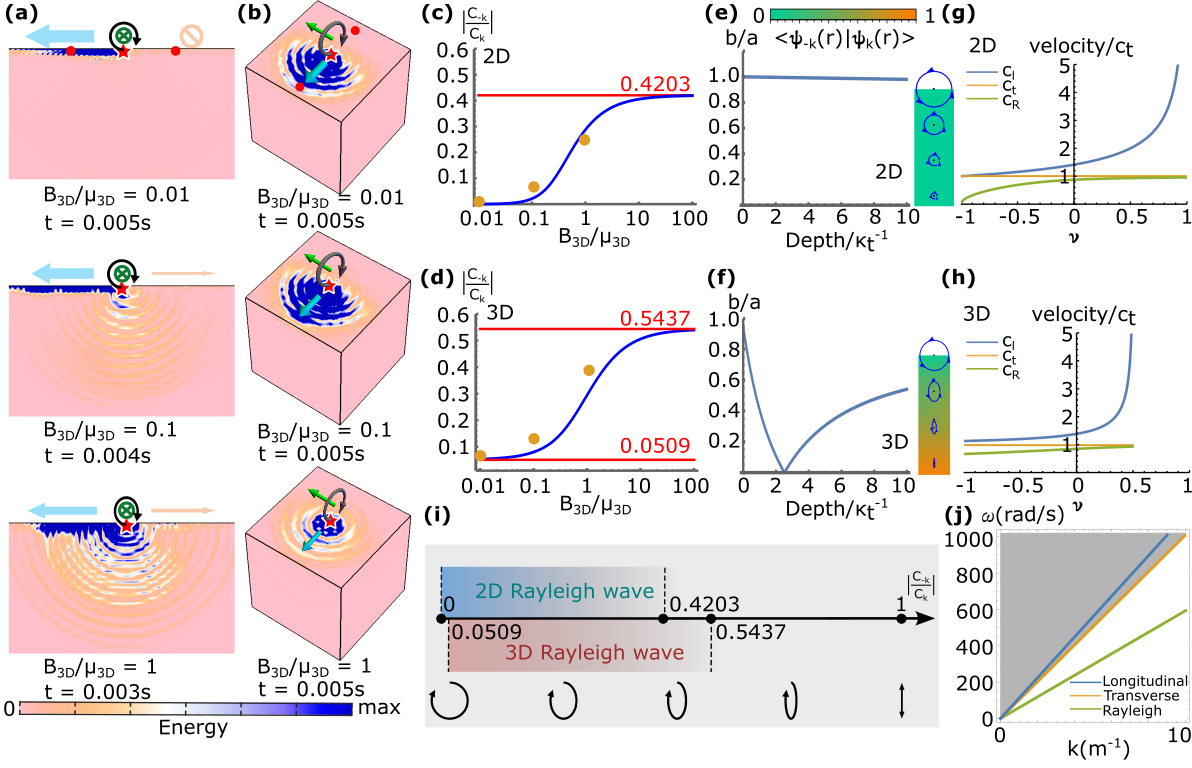}
\caption{\small 
One-way transport and spin-momentum locking of Rayleigh waves.
Simulated spatial energy distribution for 2D {\bf (a)} and 3D {\bf (b)} continua with different $B_{3D}/\mu_{3D}$ values, using a prescribed particle displacement at  boundary point $r$ following $\psi_k(r)$. 
Green arrows denote the spin direction (parallel to the surface) and blue arrows denote the strong propagation direction. The relative displacement magnitude of the red points marked in {\bf (a)} and {\bf (b)} is evaluated to measure  $|C_{-k}/C_{k}|$, shown as yellow points in {\bf (c)} and {\bf (d)}. 
The shear modulus ($\mu_{3D}$) is kept constant at $0.01$ GPa throughout all the simulations and the bulk modulus ($B_{3D}$) varied.
{\bf (c,d)} Dependence of the one-way propagation metric on the elastic moduli ratio in 2D {\bf{(c)}} and 3D {\bf{(d})}. The blue curves are analytic results, and the red lines are bounds. {\bf (e-f)} The dependence of polarization of the Rayleigh wave at various depths when $B_{3D}/\mu_{3D} = 0$ 
(normalized using the decay length $\kappa_t$ of the transverse polarized component of the edge mode),  depicted as the ratio of the short to long axes ($b/a$) of the ellipse of polarization (analytic). 
The inset illustrates the polarization (blue ellipses) and $\vec{\psi}_{-k}(r)^*\cdot \vec{\psi}_{k}(r)$  (color) of the Rayleigh waves at different depth for 2D {\bf (e)} and 3D {\bf (f)}. 
Sizes of the ellipses are proportional to the amplitude of the Rayleigh waves at the ellipses' center.
{\bf (g-h)} Dependence of the longitudinal ($C_l$),  transverse  ($C_t$), and  Rayleigh ($C_R$) wavespeeds on the Poisson's ratio ($\nu$) in 2D {\bf (g)} and 3D {\bf (h)} (analytic).
{\bf (i)} A summary of the one-way transport metric in 2D and 3D media. 
{\bf (j)} 
Dispersion of longitudinal, transverse, and Rayleigh waves in 2D continua at $B_{3D}/\mu_{3D}=0.1$ (analytic). 
The shaded area represents bulk modes. 
    }
    \label{Fig1}
\end{figure*}

In this article, we show that 100\% spin-momentum locking can be achieved using edge waves of two-dimensional (2D) isotropic continua/lattices in the limit of vanishing bulk modulus $B\to 0$ (which implies auxeticity, where Poisson's ratio $\nu=-1$). These modes are below the bulk wave bands (\textit{i.e.} lower wavespeeds), and the system doesn't need gaps between bulk bands. 
We show that the spin-momentum locking shown herein is beyond that of classical Rayleigh waves---which applies to semi-infinite bodies with flat surfaces
---and supports backscattering-free propagation (Fig.~\ref{Fig1})
under \emph{arbitrary geometries}, regardless of microscopic details (Fig.~\ref{Fig2}). 
We  verify this theory in experiments using auxetic Maxwell lattices. 
Although they have no connection to valley, spin, or mirror Chern numbers,
these waves are protected by the Kane-Lubensky index defined for Maxwell lattices~\cite{kane2014topological,lubensky2015phonons,mao2018maxwell} and continua~\cite{Sun2020} when the system is close to the $B\to 0$ limit. Furthermore, 
we show that a new topological winding number can be defined for these waves, which plays the role of discrete angular momentum eigenvalues,
generalized to arbitrary geometries, and protects the backscattering-free propagation of these waves.  

\section{Results}

\subsection{One-way transport and spin-momentum locking of Rayleigh waves}
How to excite a wave that only travels in $+k$ and not $-k$? In principle this can be achieved by applying forces on the media to exactly follow the $+k$ mode in space-time. A more interesting question, however, is how to create one-way propagation when one can only apply a periodic force $F$ on one point $r$ on the edge of the medium.  
For simplicity, we first consider the simple case where the material is a semi-infinite plane, and we will discuss general geometries next.  
This point source will generate a response in all modes at (angular) frequency $\omega$, and the amplitude of each mode $\vec{\psi}_\alpha$ will be proportional to $\vec{\psi}_\alpha(r)^* \cdot \vec{F}$, where $^*$ denotes complex conjugate (see Supplementary Materials (SM) for details).
Given that $r$ is on the edge, this overlap is much greater for edge localized modes than bulk waves.  
Furthermore, these edge modes appear in pairs of $\vec{\psi}_{k}$ and $\vec{\psi}_{-k}$, where $k$ is the momentum (wavevector) along the edge, and a generic point source $\vec{F}$ excites both $\vec{\psi}_{k}$ and $\vec{\psi}_{-k}$. 

To approach ``one-way transport''
in the $+k$ direction, one needs to maximize the ratio of the resulting amplitudes of $\vec{\psi}_{k}$ and $\vec{\psi}_{-k}$. 
This can be achieved by choosing $\vec{F} = \vec{\psi}_{k}(r)$.
A metric for asymmetric edge transport can then be defined as
\begin{equation}\label{ratiocontinuum}
\Big{\vert}\frac{C_{-k}}{C_{k}}\Big{\vert}=\Big{\vert}\frac{\vec{\psi}_{-k}(r)^* \cdot \vec{F}}{\vec{\psi}_{k}(r)^* \cdot \vec{F}}\Big{\vert}
=\Big{\vert}\frac{\vec{\psi}_{-k}(r)^*\cdot \vec{\psi}_{k}(r)}{\vec{\psi}_{k}(r)^* \cdot \vec{\psi}_{k}(r)}\Big{\vert},
\end{equation}
\noindent which vanishes in the limit of one-way transport. 
This ratio not only describes how external forces at the boundary point $r$ excite left- vs right- going waves, but also describes impurity scattering as the wave propagates in the bulk close to the edge, where impurities creates new sources $F$.  
One example where $|C_{-k}/C_{k}|=0$ 
is the quantum spin Hall effect, where spin-polarized topological edge states lead to orthogonal states for $k$ and $-k$.

Here, we consider instead Rayleigh waves, 
\begin{equation}\label{ctsmodeshape}
\vec{\psi}_{\pm k}(x,y)=e^{i(\pm kx-\omega t)}\left(\frac{1}{2}\xi\sqrt{1+\frac{\mu}{B}},\pm\frac{1}{4}i\xi(2-\xi^2)\sqrt{\frac{1+(\mu/B)}{1-\xi^2}}\right)^T
\end{equation}
at the edge ($y=0$) of a semi-infinite plane, 
and $\xi=\xi(B/\mu)$ is the ratio of Rayleigh to transverse wave speeds and a function of the ratio of bulk modulus $B$ and shear modulus $\mu$ (see SM for details). 
This pair of waves are linear combinations of longitudinal and transverse waves, satisfying stress-free boundary conditions at $y=0$, and their longitudinal and transverse components decay into the bulk with different decay lengths \cite{Rayleigh1887}. 
Here $B$ and $\mu$ are moduli for 2D isotropic continuum elasticity.  To make this discussion relevant for experiments, we relate them to 3D elastic moduli (subscripts 3D) of the material that consists this 2D plate (under plane stress condition), which reads 
$B=3B_{3D}\mu_{3D}/(B_{3D}+\frac{4}{3}\mu_{3D})$ and $\mu=\mu_{3D}$, and use 3D moduli in our computational results.

Interestingly, at the limit of $B/\mu \to 0$, 
(equivalently $\xi\to 0$), this pair of waves become
\begin{equation}\label{analyticalmodeshape}
\vec{\psi}_{\pm k}(x,y)=\frac{1}{\sqrt{2}}e^{i(\pm kx-\omega t)}\left(1,\pm i\right)^T,
\end{equation}
which are circularly polarized. 
In this case, $|C_{-k}/C_{k}|=0$,
leading to $100\%$ one-way transport.  
The general case of 
$B/\mu>0$ leads to ellipsoidal polarization and $|C_{-k}/C_{k}|>0$.  
In the limit of $B/\mu\to \infty$, $|C_{-k}/C_{k}|\to 0.4203<1$, meaning that the propagation along $\pm k$ is still biased (Fig.~\ref{Fig1}(a,c) with simulation described in Methods). 

The circular polarization of the Rayleigh waves in the limit of $B/\mu\to 0$ suggests that a spin-momentum locking mechanism can be defined for this one-way transport.  
Indeed, similar to the spin angular momentum of photons, where photons of $\pm 1$ spins correspond to quanta of right (R) and left (L) polarized electromagnetic waves, a spin angular momentum can be defined for phonons~\cite{francesco2012conformal}.  
In the SM, we show that the rotational symmetry of the media leads to conserved angular momentum density, following Neother's theorem, and this angular momentum $\vec{j} = \vec{l} +\vec{s}$ can be expressed as a sum of an orbital ($l$) and spin ($s$) angular momentum density of the wave, where
\begin{align}\label{AM}
\vec{l} &= \vec{r}\times\vec{p}, \quad
\vec{s} = \rho \vec{u}\times(\partial_t \vec{u}),
\end{align}
$\rho$ is the mass density, and $\vec{p}=-\rho \partial_t u_i \vec{\grad} u_i$ is the density of linear momentum.  
Following this, the Rayleigh waves carry physical spin angular momentum pointing along $\hat{z}$ (normal to the 2D plane), and under standard canonical quantization, the spins of these phonons take $\pm 1$.  

What about 3D Rayleigh waves?  
A similar calculation can be done, giving a pair of Rayleigh waves for any propagation direction on the 2D surface of a semi-infinite 3D solid (see SM for details)~\cite{Wavepropauxetic,lipsett1988reexamination}.
However, the asymmetric edge transport ratio never reaches zero ($|C_{-k}/C_{k}|>0.0509$, Fig.~\ref{Fig1}(b,d)).  Thus, $100\%$ one-way transport, in the manner defined herein using Rayleigh waves, is only achieved for 2D solids. 
More importantly, 3D Rayleigh waves are far from ``perfect'' in terms of spin-momentum locking, even in the $B\to 0$ limit. 
This is because in a 3D solid, although the Rayleigh-wave mode shape is close to circular polarization at the surface layer,
as one moves into the bulk, the Rayleigh-wave mode shape quickly becomes elliptical~\cite{lipsett1988reexamination} (Fig.~\ref{Fig1}(f)), 
i.e., $|C_{-k}/C_{k}|>0$ in the bulk, and thus any impurity would result in backscattering.
In contrast, 2D Rayleigh waves have $|C_{-k}/C_{k}|=0$ at all depths (Fig.~\ref{Fig1}(e)), making it immune to backscattering.  

\begin{figure*}[t]
    \centering
    \includegraphics[width=\linewidth]
    {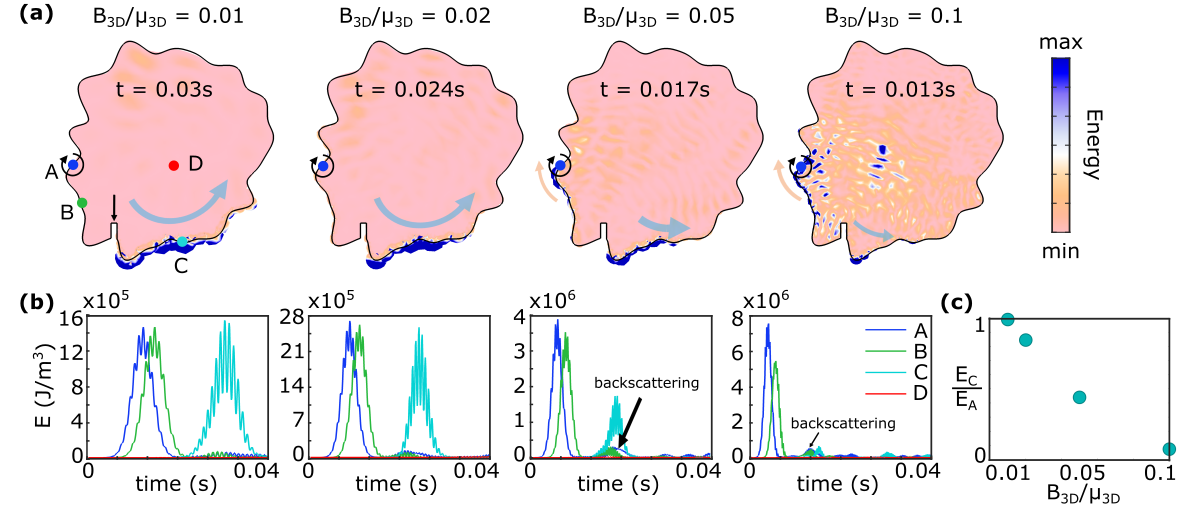}
    \caption{\small Backscattering-free propagation of one-way edge waves in arbitrary geometries. 
    {\bf (a)} Simulated energy distribution for 2D continua  
    with different values of $B_{3D}/\mu_{3D}$: $0.01, 0.02, 0.05,$ and $0.1$ (at different times). The clockwise actuation (Gaussian wave packet) excites right-going edge waves.
    To characterize the backscattering, a sharp defect is created at the boundary (black arrow). {\bf (b)} 
    Energy of points $A$ (actuation point), $B$, $C$, and $D$ (each marked in {\bf (a)}) as functions of time. {\bf (c)} Energy ratio between points $C$ and $A$ (cumulative over time) at different $B_{3D}/\mu_{3D}$ demonstrates vanishing backscattering in the limit of $B/\mu\to 0$. 
    }
    \label{Fig2}
\end{figure*}

\subsection{Backscattering-free one-way transport in arbitrary geometries}
The Rayleigh wave formulation discussed above offers a concise picture to characterize how spin-momentum locking arises in these edge waves in the simple geometry of semi-infinite media.  
However, the observed one-way transport appears to be robust in materials with very irregular geometries with no backscattering in the limit of $B/\mu\to 0$ (Fig.~\ref{Fig2}).  
To reveal the physical origin of this phenomenon, we propose a new formulation based on mapping of zero modes in the $B/\mu\to 0$ limit in 2D to complex analytic functions and perturbatively extending them to low-frequency edge waves at small $B/\mu$.  
By doing this, we show that such spin-momentum locking is a general phenomenon in 2D, small $B/\mu$ materials of any shape, and these edge waves are characterized by a new topological winding number that plays the role of discrete angular momentum eigenvalues despite the arbitrary geometry.

We first use the complex plane to rewrite the 2D elasticity problem, by defining coordinate $z=x+iy$ and displacement field $\psi=u_x + i u_y$.   
In the limit of $B/\mu\to 0$, the wave equation and boundary conditions can then be written as  (see SM for details)
\begin{equation}\label{complex}
\begin{aligned}
\rho\frac{\partial^2\psi(z,z^*,t)}{\partial t^2}&=4\mu \frac{\partial^2\psi(z,z^*,t)}{\partial z \partial z^*}, \quad z\in \Omega,\\
\frac{\partial \psi(z,z^*,t)}{\partial z^*}&=0, \quad z\in \partial\Omega, 
\end{aligned}
\end{equation}
where $\Omega$ and $\partial\Omega$ represent the bulk and edge of the system, respectively.
Solutions of Eq.~\eqref{complex} can be obtained via separation of variables $\psi_{\pm} (z,z^*,t) = \phi(z,z^*) \, e^{\pm i\omega t}$. 
Because $e^{+ i\omega t} $ rotates the displacement vector counterclockwise and $e^{- i\omega t} $ rotates it  clockwise over time, these two solutions correspond to spin up and down states.  
This equation has $\omega=0$ solutions $\phi_0$ given by $\partial \phi_0 / \partial z^*=0$, \textit{i.e.}, arbitrary bounded analytic functions $\phi_0(z)$.  
These solutions as static zero modes have been characterized in Ref.~\cite{sun2012surface}, and they can be derived from the fact that zero modes in 2D materials at $B/\mu= 0$ are conformal transformations, which map to analytic functions~\cite{sun2012surface,czajkowski2022conformal}.  

Here we are interested in the dynamics of waves at small but nonzero $B/\mu$, which can be solved perturbatively from the zero modes.  
Same as the perturbation theory for the Schr\"{o}dinger equation~\cite{griffiths2017introduction}, to the leading order, small perturbations only modify the frequencies (eigenvalues), while wavefunctions remain analytic functions. Thus, at small $B$, these zero modes acquire finite frequencies $\omega\propto\sqrt{B/\mu}$, 
\begin{equation}\label{zeroth}
\psi_{\pm} (z,z^*,t) = \phi_0(z) \, e^{\pm i\omega t}.
\end{equation}

\begin{figure*}[t]
    \centering
    \includegraphics[width=\linewidth]
    {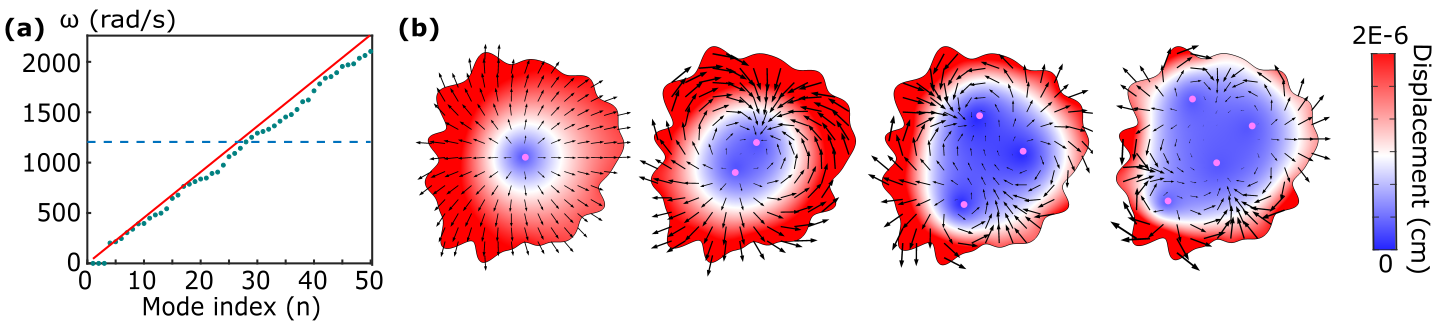}
    \caption{Eigenfrequencies and phase-winding of one-way edge waves in arbitrary geometries. 
    {\bf (a)} Computed (green dots) and analytic (red line, following Eq.~\eqref{FreqN})
    eigenfrequencies of a 2D media of irregular shape plotted against the mode index at $B_{3D}/\mu_{3D} = 0.01$. 
    Modes above the blue dashed line are bulk modes.  {\bf (b)} Profiles of modes $n=4,6,8,10$ (corresponding to winding numbers $N_0=1,2,3,4$)
    as examples of waves described by Eq.~\eqref{zeroth}. 
    Pink dots mark $\phi=0$ points in the bulk, the number of which equals the winding number $N_0$ (Eq.~\eqref{argumentprinciple}). 
    The black arrows represent the displacement vectors, which rotate $N_0$ times around the boundary, and vanish at $N_0$ points in the bulk. 
    }
    \label{Fig3}
\end{figure*}

How do these waves propagate in space and time? 
The irregular geometry we consider here exhibits no translational or rotational symmetries, so linear momentum $k$ and orbital angular momentum $l$ are no longer conserved quantities.  
Instead, a winding number can be defined by applying the argument principle on the analytic functions $\phi_0$, which we suggest plays the role of ``quantized'' orbital angular momentum eigenvalues,
\begin{equation}\label{argumentprinciple}
\frac{1}{2\pi i}\int_{\partial\Omega}d\ln(\phi_{0}(z))=\frac{1}{2\pi i}\int_{\partial\Omega}\frac{\phi'_{0}(z)}{\phi_{0}(z)}dz=N_0-N_{\infty},
\end{equation}
where $N_0$ and $N_{\infty}$ are the numbers of zeros and poles of $\phi(z)$ inside the bounded domain $\Omega$. The function $\phi_0(z)$ being bounded in $\Omega$ implies $N_{\infty}=0$.  
On the other hand, edge localized modes $\phi_0(z)$ must have zeros in $\Omega$ (Rouché's theorem, as detailed in SM), so $N_0 \ge 1$.  
This indicates that all zero modes $\phi_0$ wind its phase by at least one cycle of $2\pi$ as one goes along the boundary counterclockwise.  
For a finite 2D material at $B/\mu=0$, there are infinitely many such zero modes, and they each exhibit a winding number $N_0$, which is also the number of points in the bulk where the amplitude of the zero mode reaches zero (counting multiplicity), as shown in Fig.~\ref{Fig3}(b).  
In addition, zero modes with $N_0$ zeros can be expanded as analytic functions as 
\begin{equation}
\phi_{0,n}(z)=a_{N_0} z^{N_0}+...+a_1z+a_0 .
\end{equation}
It is thus clear that zero modes with larger $N_0$ are more localized on the edge, which is consistent with predictions from  conformal symmetry~\cite{sun2012surface}.

This immediately leads to spin-momentum locking at small $B/\mu$:
constant phase points of $\psi_{+} (z,t) = \phi_0(z) \, e^{+ i\omega t}$ move clockwise and constant phase points of $\psi_{-} (z,t) = \phi_0(z) \, e^{- i\omega t}$ move counterclockwise in time. 
Therefore, this winding number is the extension of discrete angular momentum eigenvalues from perfect circular geometry to arbitrary geometries in the limit of $B/\mu\to 0$.  
It is worth noting that such an extension is not generally defined for any 2D material, where wave eigenstates are usually affected by irregular shapes of the boundary and do not exhibit a clear propagation direction or momentum (see examples in Fig.~\ref{Fig2}). 
It is the $B/\mu\to 0$ limit that permits such a clean definition of an orbital angular momentum.

The two waves $\psi_{\pm} (z,t)$ are orthogonal, satisfying the
one-way transport criterion $|C_{-k}/C_{k}|=0$ defined in Eq.~\eqref{ratiocontinuum}.
This can be seen by returning from the complex plane notation to the conventional wave notation with $x,y$ components, where
\begin{equation}\label{phipolarization}
\phi_0(z) \, e^{\pm i\omega t} \to \Re{\phi_0 (x+iy) e^{i\omega t}  (1,\pm i)^{T}},
\end{equation}
which are circularly polarized, with left- (L) and right- (R) waves orthogonal to one another.  
This orthogonality is immune to shape irregularities of the boundary, as it stems from exact zero modes in the $B/\mu\to 0$ limit, and does not rely on any notion of plane waves or Rayleigh waves of semi-infinite geometry.  
Thus, these circularly polarized waves approach backscattering-free as $B/\mu\to 0$ for any geometry (Fig.~\ref{Fig2}).

The spectrum of these edge modes also resemble waves characterized by discrete angular momentum. 
For 2D materials with shapes of no particular symmetry, to leading order in $B/\mu$, these modes have frequencies 
\begin{equation}\label{FreqN}
\omega_n \simeq \frac{2\pi N_{0}(n) v_{phase}}{L}
\simeq \frac{\pi n v_{phase}}{L}
\end{equation}
where $n$ labels the modes, $L$ is the perimeter of the bounded domain $\Omega$, $v_{phase}=\xi(B/\mu)\sqrt{\mu/\rho}$ is the phase velocity of  Rayleigh waves of the given material. 
The winding number is related to the mode index by $N_0(n)=\lceil (n-3)/2 \rceil$ where $\lceil \cdot \rceil$ is the ceiling function (note that each zero mode at $B=0$ generates a pair of edge modes at $B>0$, and there are three rigid body modes that remain at $\omega=0$), and this reduces to $N_0(n)\simeq n/2$ at large $n$, 
as shown in Fig.~\ref{Fig3}(a).  
Frequencies of these edge modes extend to the point of the first bulk mode, which is at the scale of 
$\sqrt{\mu/\rho}/L$, noting that the longitudinal and transverse wavespeeds converge as $B/\mu\to 0$ (see Fig.~\ref{Fig1}(g)). 

\begin{figure*}[t]
    \centering
    \includegraphics[width=1\linewidth]{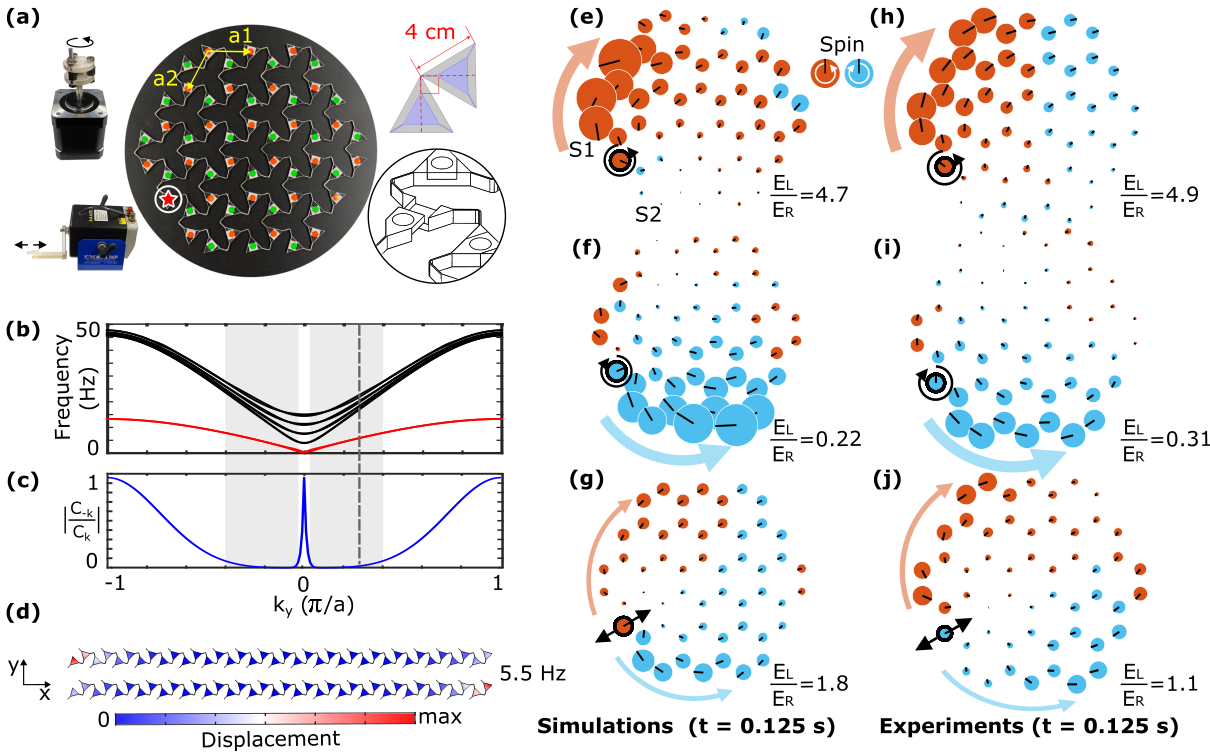}
    \caption{\small 
    Experimental verification of spin-momentum locking and one-way edge waves in an auxetic Maxwell lattice.
    {\bf (a)} 
    The self-dual twisted kagome lattice~\cite{fruchart2020dualities}. The red star marks the actuation point. 
    Red and green solid circles denote the triangles' centroids, which are used for image tracking.
    The 2D lattice has primitive basis vectors $(\vec{a}_1, \vec{a}_2)$ with a lattice constant of $a = 5.71$ cm.
    Insets: (left) circular actuator (with a radius of $5$ mm), and linear actuator; and (right) unit-cell geometry (gray: triangles with ideal pin-joint; blue: triangles with beam-like elastic connectors) and assembly diagram. 
    {\bf (b)} 
    Super-cell band structures simulated in FEM for the elastic lattice in {\bf (a)} with edge bands marked in red.
    {\bf (c)} 
    $|{C_{-k}}/{C_{k}}|$ of the lowest edge band. 
    Shaded area indicates $|{C_{-k}}/{C_{k}}|<0.1$. 
    Dashed line marks $k = 0.28 {\pi}/{a}$.
    {\bf (d)} Two simulated edge mode profiles at $k = 0.28 {\pi}/{a}$, also showing the geometry simulated in {\bf (b)}.
    {\bf (e-g)} Time domain FEM simulations and {\bf (h-j)} experimental results. 
    Black circles and arrows indicate the location and the direction of the excitation, respectively. Solid circle sizes are proportional to velocity magnitudes and black lines indicate velocity vector directions. Orange (light blue) color indicates the sign of the rotation of the triangle centroids' velocity vector, corresponding to counterclockwise (clockwise) spins.
    ${E_L}/{E_R}$ indicates the energy ratio between left (L) and right (R) propagating modes. 
    }
    \label{Fig4}
\end{figure*}

\subsection{One-way transport in twisted kagome lattices}

To demonstrate this spin-dependent edge transport experimentally, we choose the twisted kagome lattice, which is known to exhibit $B/\mu=0$ in the ideal pin-joint limit (no bending stiffness between triangles)~\cite{sun2012surface}.  
The lattice we consider herein (Fig.~\ref{Fig4}(a)) deviates from this limit, as we use beam-like elastic connectors instead of pin joints. We measure the Poisson's ratio of this lattice to be $\nu\simeq-0.488$ (see SM), corresponding to $B/\mu=0.344$.

Pairs of branches of exponentially localized edge zero modes are known to arise in twisted kagome lattices under mixed (periodic horizontal and open vertical) boundary conditions, at opposite open edges~\cite{sun2012surface}. 
They show up as the red branch in (Fig.~\ref{Fig4}(b)) via FEM super-cell analysis of the elastic kagome lattice, which are 
actually two nearly overlapping branches
At wavelengths approaching the width of the simulated domain (Fig.~\ref{Fig4}(d))
or low $k$, the two localized modes generally hybridize into Lamb-like modes \cite{lamb1917waves} (see SM). The effect of this hybridyzation can  be seen in Fig.~\ref{Fig4}(c), where $|{C_{-k}}/{C_{k}}|$ exhibit a peak near $k\to 0$.
An important exception to this is the $B/\mu\to 0$ case, wherein the two edge modes are exactly orthogonal and thus, do not hybridize and retain their asymmetric behavior. 
When $k=0.28\pi/a$, the decay length is short
(about $1.4$ unit cells) and the hybridyzation is weak, so we choose the frequency of this pair of modes 
($5.5$ Hz, Fig.~\ref{Fig4}(d)) for FEM and experiment of edge waves.

We verify spin-momentum locking and one-way edge wave propagation by actuating at the centroid of one triangle at the edge. With circularly polarized excitation, we observe spin-dependent one-way edge wave propagation in both simulation (Fig.~\ref{Fig4}(e,f)) and experiment (Fig.~\ref{Fig4}(h,i)). 
With linear actuation, we observe nearly symmetric edge-wave propagation, wherein the wave's spin is matched to the direction of propagation (Fig.~\ref{Fig4}(g,j)).
For each excitation case, we also calculate the energy ratio ${E_L}/{E_R}$ (Fig.~\ref{Fig4}(e-j)) between L and R propagating edge modes. The energy for each mode is calculated by integrating the average velocity squared (over $0.01-0.15$ s) of triangles on the edge to the left and right of the excitation point (see SM for specific triangles). Both the spatial response and the ${E_L}/{E_R}$ ratio shows a close agreement between simulation and experiment (Fig.~\ref{Fig4}). 

\section{Discussion}
We have reported a new type of one-way and backscattering-free edge wave based on spin-momentum locking in 2D auxetic materials with moduli $B/\mu \to 0$. These modes are protected by a new topological winding number that generalizes discrete angular momentum eigenvalues to arbitrary geometries.  
Unlike other topologically protected edge modes, these protected edge modes are not required to reside in a gaps between bulk bands and instead exist for a broad range of frequencies. 
This is important, because, compared to the  bandgap-tied mechanisms (quantum Hall, quantum spin, valley, or mirror Hall types): i) it opens previously inaccessible ranges of low frequencies for topologically protected transport, 
and ii) it does not necessitate any precise periodic or locally-resonant structures, and therefore expands the range of usable frequencies in a given material, simplifies fabrication, and minimizes dissipation. 

As per its broadband functionality, the one-way edge mode transport mechanism shown herein is applicable to essentially any auxetic lattice or continuum in the $B/\mu \to 0$ limit, regardless of microscopic details. 
As auxetic materials have been studied for decades 
\cite{saxena2016three}, 
there exists a broad potential set of candidate materials, including top-down manufactured structural materials, such as 3D printed re-entrant honeycomb lattices 
\cite{yang2015mechanical}, and processed bottom-up manufactured materials, such as thermomechanically compressed polyurethane foams \cite{li2016successful}. 
Some naturally occurring auxetic materials have also been found, exhibiting anisotropic negative Poisson's ratios (albeit still far from $\nu\to-1$), such as $\alpha$-cristobalite and some metals \cite{greaves2011poisson}. Our theory predicts that asymmetric edge transport exists in these materials, although perfect backscattering-free one-way transport calls for proximity to $\nu\to-1$ and low dissipation. 

Finally, we emphasize that the ``spin'' of these edge waves are real, in the sense that they are directly connected to angular momentum, instead of the ``pseudo spins'' typically unitized to introduce the spin Hall effect in phononic systems~\cite{susstrunk2015observation, mousavi2015topologically, miniaci2018experimental}. Because of this physical spin, it has been previously suggested that such circularly polarized waves may be able to couple to spin degrees of freedom of other signals \cite{long2018intrinsic} such as photons \cite{temnov2012ultrafast, guddala2021topological,choi2022chiral} and magnons \cite{holanda2018detecting}. Similarly, we speculate that such coupling and topological protection may find use in new types of optomechanical device strategies \cite{he2016optomechanical}. 

\section{Methods}

\subsection{Maxwell lattice manufacturing}

The self-dual kagome lattice with $90^{\circ}$ rotated solid equilateral triangles connected via long and slender beams shown in Fig.~\ref{Fig4}(a) was designed to be assembled together from repeating sub-lattice elements. The sub-lattice elements (design shown in SM) were machined from $6.2$ mm thick polycarbonate sheets via desktop CNC mill then assembled together via press fits. 

\subsection{Experimental edge wave characterization}

A misaligned shaft coupling ($5$ mm central offset) is installed on a stepper motor to induce a circular displacement path to the actuated triangle. The motor is controlled by a RAMBo board via MATLAB R2022b. A ball bearing is used at the end of the shaft to minimize rotation of the actuated triangle. The excitation frequency is $5.5$ Hz. An electrodynamic exciter (The Modal Shop, Smartshaker, K2007E01) is driven sinusoidally at $5.5$ Hz by a signal generator (Tektronix, AFG3022C) to perform the linear actuation. A digital camera records the displacements of the lattice under actuation from above at a sampling rate of $240$ Hz. The locations of the red and green markers in Fig.~\ref{Fig4}(a) are tracked from the recorded image sequence by using the \textit{imfindcircles} command in MATLAB R2022b. The spatial measurement resolution is calculated to be $0.37$ mm for the circular actuation cases, and $0.48$ mm for the linear actuation cases. The location coordinates are smoothed in the time domain with a Gaussian-weighted moving average filter using the \textit{smoothdata} command with a window size of $20$ in MATLAB R2022b.

\subsection{FEM simulation}

The FEM simulations are performed using the COMSOL Multiphysics Structural Mechanics Module. 
The 2D and 3D continua depicted in Figs.~\ref{Fig1}, \ref{Fig2}, and \ref{Fig3} use user-defined material properties with a mass density of $960$ kg/m$^3$ and a shear modulus of $0.01$ GPa, and differing bulk moduli, with extra fine mesh (Fig.~\ref{Fig1}: Mapped; Figs.~\ref{Fig2},\ref{Fig3}: Free Triangular, minimum element size $0.0075$ cm and maximum element size $2$ cm). The plane stress approximation has been employed in the 2D simulations. The dimensions of these systems are approximately $1$ m $\times~1$ m for 2D ($1$ m $\times~1$ m $\times~1$ m for 3D).
Figures~\ref{Fig1} and \ref{Fig2} are simulated in the time domain, with a time step of $0.0001$ s, and the simulations are carried out from $0$ s to $0.05$ s. Figure~\ref{Fig3} is simulated in the frequency domain.
The governing equation is given by $\rho \frac{d^2u}{dt^2} = \nabla \sigma + \vec{F(r,t)}$, where $\vec{F(r,t)}$ represents the time-dependent excitation at a boundary point, expressed in the form of $\vec{\psi}_{k}(r,t)$ (noting that a prescribed displacement is applied). 
In the 2D continua, the excitation spin direction is out-of-plane, while in the 3D continua, it is parallel to the surface.
In Fig.~\ref{Fig1}, a free boundary condition is applied to the boundary with excitation, and low-reflecting boundary conditions are employed on the remaining boundaries. This setup allows us to evaluate the ratio $|C_{-k}/C_{k}|$. This ratio is calculated by taking the ratio of the square root of the energy at the marked points, which are located on opposite sides of the excitation, over an extended period of time.
In Figs.~\ref{Fig2} and \ref{Fig3}, free boundary conditions are applied. 
The excitation frequency is adjusted to $1000$ Hz for different $B_{3D}/\mu_{3D}$ in 3D. 
The excitation frequency varies depending on the value of $B_{3D}/\mu_{3D}$ in 2D, with frequencies of $500$ Hz, $1500$ Hz, and $2000$ Hz used for $B_{3D}/\mu_{3D} = 0.01$, $B_{3D}/\mu_{3D} = 0.1$, and $B_{3D}/\mu_{3D} = 1$, respectively. The simulated results are smoothed in the time domain following the same process as the experimental data.

The Maxwell lattices shown in Fig.~\ref{Fig4} are filled with polycarbonate material, which has a mass density of $1190$ kg/m$^3$, a Young's modulus of $0.751$ GPa, and a Poisson's ratio of $0.3182$. 
In the super-cell analysis, periodic boundary conditions are applied along $y$ direction to conduct parametric sweeps of wave vectors.
The ratio $|{C_{-k}}/{C_{k}}|$ in Fig.~\ref{Fig4} {\bf (c)} is computed using equation \ref{ratiocontinuum}, utilizing the eigenvectors (displacement) $\vec{\psi}{k}(r)$ and $\vec{\psi}{-k}(r)$ obtained for the first band. Here, $r$ represents the centroid of the first left triangle, as the energy of the first band localizes at the left edge when $k$ is not in the small limit.
In Fig.\ref{Fig4} {\bf (e-g)}, the excitation is applied at a single triangle centroid on the boundary, with a frequency $f = \omega/(2\pi) = 5.5$ Hz. 
This frequency corresponds to the lowest edge mode frequency in the super-cell analysis when $k =\pm 0.28\pi/a$, where $a$ is the lattice constant.

\section*{Acknowledgement}
W. C., N. C., K. S., and X. M. acknowledge support from the Office of Naval Research (MURI N00014-20-1-2479). 
K. Q., N. B., and X. M. acknowledge support from the US Army Research Office (Grant No. W911NF-20-2-0182).\\

\section*{Data and code availability}
The data and code that support the findings of this study are available from the corresponding authors on reasonable request.\\

\section*{Author contributions}
W. C. performed the full-wave simulations, K. Q. \ performed the fabrication, and experimental characterization. 
N. C., X. M. and K. S. \ developed the mathematical framework. 
W. C. and K. Q. analyzed and interpreted the data. All authors wrote the manuscript. N. B., X. M., and K. S. \ supervised the project.\\

\section*{Competing interests}
The authors declare no competing interests.\\

\section*{Supplemental Material}
The supplemental materials contain detailed mathematical discussion on 2D and 3D Rayleigh wave mode shape, criterion for perfectness of Rayleigh wave one-way transport in continuum systems and discrete lattices, the spin and momentum of elastic waves as well as full spin and momentum locking of Rayleigh wave in 2D material with $B/\mu\ll 1$.\\

\bibliography{references.bib}

\clearpage
\begin{center}
\begin{widetext}
\textbf{\large Backscattering-free edge states below all bands in two-dimensional auxetic media: Supplementary Materials}
\end{widetext}
\end{center}
\setcounter{equation}{0}
\setcounter{figure}{0}
\setcounter{table}{0}
\setcounter{page}{1}
\makeatletter
\renewcommand{\theequation}{S\arabic{equation}}
\renewcommand{\thefigure}{S\arabic{figure}}
\renewcommand{\bibnumfmt}[1]{[S#1]}
\renewcommand{\citenumfont}[1]{S#1}

\section{3D Rayleigh Waves}

Here we review the linear elastic wave equation and Rayleigh wave formula in 3D. The main reference of this formulation is Ref.~\cite{landau1986theory}. Consider a homogeneous elastic material in $x^3\leq 0$ region in the $x^1$-$x^2$-$x^3$ space with bulk modulus $B_{3D}$, Lamé constant $\mu_{3D}$ and $\lambda_{3D}=B_{3D}-\frac{2}{3}\mu_{3D}$. 
Let $\vec{v}$ be the displacement vector field. The strain tensor is given by 
\begin{equation}\label{strain}
u_{jk}=\frac{1}{2}(\delta_{ik}\frac{\partial v^{i}}{\partial x^j}+\delta_{ij}\frac{\partial v^{i}}{\partial x^k}). 
\end{equation}
The elastic energy density for homogeneous material is
\begin{equation}\label{energydensity}
\epsilon=\frac{1}{2}\lambda_{3D} \delta^{ij}\delta^{kl}u_{ij}u_{kl}+\mu_{3D} \delta^{ij}\delta^{kl}u_{ik}u_{jl}
\end{equation}
The stress tensor $\sigma^{ij}$ is defined as
\begin{equation}
d\epsilon=\sigma^{ij}du_{ij},
\end{equation}
From this we obtain the stress tensor
\begin{equation}
\sigma^{ij}=\lambda_{3D} \delta^{ij}\frac{\partial v^a}{\partial x^a}+\mu_{3D}\delta^{ib}\frac{\partial v^j}{\partial x^b}+\mu_{3D}\delta^{jc}\frac{\partial v^i}{\partial x^c}.
\end{equation}
Let $\rho$ be the mass density of the elastic material. 
The equation of motion is followed by Newton's second law
\begin{equation}
\rho\frac{\partial^2 v^i}{\partial t^2}=\frac{\partial \sigma^{ij}}{\partial x^j},
\end{equation}
which is equivalent to
\begin{equation}\label{ctseom3D}
\rho\frac{\partial^2 \vec{v}}{\partial t^2}=(\lambda_{3D}+\mu_{3D})\vec{\nabla}\vec{\nabla}\cdot\vec{v}+\mu_{3D} \nabla^2\vec{v}.
\end{equation}
From Helmholtz decomposition, we can write
\begin{equation}\label{decomp}
\vec{v}=\vec{v_l}+\vec{v_t},
\end{equation}
where $\vec{v_l}$ is the curl free and $\vec{v_t}$ is divergence free. 
By taking curl on both sides of \eqref{ctseom3D}, we get
\begin{equation}
\vec{\nabla}\times(\rho\frac{\partial^2 \vec{v_t}}{\partial t^2}-\mu_{3D} \nabla^2\vec{v_t})=0.
\end{equation}
Since $\vec{v_t}$ is divergence free, the vector field $\rho\frac{\partial^2 \vec{v_t}}{\partial t^2}-\mu_{3D} \nabla^2\vec{v_t}$ is both gradient free and curl free, hence
\begin{equation}\label{longeom3D}
\rho\frac{\partial^2 \vec{v_t}}{\partial t^2}-\mu_{3D} \nabla^2\vec{v_t}=0.
\end{equation}
Similarly, by taking gradient on \eqref{ctseom3D}, we have 
\begin{equation}\label{transeom3D}
\rho\frac{\partial^2 \vec{v_l}}{\partial t^2}-(2\mu_{3D}+\lambda_{3D}) \nabla^2\vec{v_l}=0.
\end{equation}\\
Without loss of generality, we may assume the surface wave propagates in the $x^1$ direction. Let the longitudinal and transverse components of the Rayleigh wave be 
\begin{equation}\label{rayleighansatz3d}
\begin{aligned} 
&\vec{v_l}=e^{i(kx^1-\omega t)+\kappa_l x^3}(\alpha^1, \alpha^2,\alpha^3)^T,\\
&\vec{v_t}=e^{i(kx^1-\omega t)+\kappa_t x^3}(\beta^1, \beta^2,\beta^3)^T.
\end{aligned}
\end{equation}
Plugging \eqref{rayleighansatz3d} into \eqref{longeom3D}, \eqref{transeom3D}, we have
\begin{equation}
\kappa_l=\sqrt{k^2-\frac{\omega^2}{c_l^2}}, \; \kappa_t=\sqrt{k^2-\frac{\omega^2}{c_t^2}},
\end{equation}
where $c_l=\sqrt{\frac{\lambda_{3D}+2\mu_{3D}}{\rho}}$ and $c_t=\sqrt{\frac{\mu_{3D}}{\rho}}$ are the speed of longitudinal wave and transverse wave. 
$x^3=0$ surface being an open boundary gives the boundary conditions
\begin{equation}\label{ctsbdy3d}
\begin{aligned}
&\sigma^{13}=\mu_{3D}(\frac{\partial v^3}{\partial x^1}+\frac{\partial v^1}{\partial x^3})=0,\\
&\sigma^{23}=\mu_{3D}(\frac{\partial v^3}{\partial x^2}+\frac{\partial v^2}{\partial x^3})=0,\\
&\sigma^{33}=\lambda_{3D}\frac{\partial v^1}{\partial x^1}+\lambda_{3D}\frac{\partial v^2}{\partial x^2}+(\lambda_{3D}+2\mu_{3D})\frac{\partial v^3}{\partial x^3}=0.
\end{aligned}
\end{equation}
The conditions $\vec{\nabla}\times\vec{v_l}=0$, $\vec{\nabla}\cdot\vec{v_t}=0$ and $\sigma^{23}=0$ give
\begin{equation}
\begin{aligned}
-&\kappa_l \alpha^1+ik \alpha^3=0,\\
& ik\beta^1+\kappa_t \beta^3=0\\
&\alpha^2=\beta^2=0
\end{aligned}
\end{equation}
which says there exists $a$ and $b$ such that
\begin{equation}\label{3dsimplicification1}
\begin{aligned}
&\alpha^1=ka, \alpha^3=-i\kappa_l a,\\
&\beta^1=\kappa_t b, \beta^3=-ikb.
\end{aligned}
\end{equation} 
Combining \eqref{rayleighansatz3d}, \eqref{ctsbdy3d} and \eqref{3dsimplicification1} we get
\begin{equation}\label{3dlinearequation}
\begin{aligned}
& b(k^2+\kappa_t^2)+2ak\kappa_l=0,\\
&2b\kappa_tk+a(k^2+\kappa_t^2)=0.
\end{aligned}
\end{equation}
In order that \eqref{3dlinearequation} has a non-zero solution, the coefficient matrix need to be of determinant zero, which gives
\begin{equation}
(k^2+\kappa_t^2)^2=4k^2\kappa_t\kappa_l.
\end{equation}
Let $\omega=c_tk\xi$, we obtain
\begin{equation}\label{3dmaster}
\xi^6-8\xi^4+8\xi^2(3-\frac{6}{3h_{3D}+4})-16(1-\frac{3}{3h_{3D}+4})=0,
\end{equation}
where $h_{3D}=\frac{B_{3D}}{\mu_{3D}}$. The corresponding modes at $x^3=0$ surface are
\begin{widetext}
\begin{equation}\label{ctsmodeshape3D}
\begin{aligned}
&\psi_{k}(x^1,x^2,0)=e^{i(kx^1-\omega t)}(-\frac{\xi^2\sqrt{1-\xi^2}}{2-\xi^2}, 0, i(\frac{2\sqrt{1-\xi^2}}{2-\xi^2}\sqrt{1-\frac{3\xi^2}{3h_{3D}+4}}-1))^T,\\
&\psi_{-k}(x^1,x^2,0)=e^{i(-kx^1-\omega t)}(-\frac{\xi^2\sqrt{1-\xi^2}}{2-\xi^2}, 0, -i(\frac{2\sqrt{1-\xi^2}}{2-\xi^2}\sqrt{1-\frac{3\xi^2}{3h_{3D}+4}}-1))^T.
\end{aligned}
\end{equation}
\end{widetext}

\section{2D Rayleigh Waves}\label{SEC:2DRayleigh}
Here we review
the elastic wave equation and Rayleigh wave formula for a flat thin homogeneous elastic sheet within a plane stress condition, with bulk modulus $B_{3D}$ and Lamé constant $\mu_{3D}$ and $\lambda_{3D}=B_{3D}-\frac{2}{3}\mu_{3D}$. 
We choose a coordinate system such that the $x^1$ and $x^2$ directions lie in the plane of the sheet and $x^3$ direction is normal to the plane. 
On the top and bottom surfaces of the sheet, we have
\begin{equation}\label{planestress}
\sigma^{13}=\sigma^{23}=\sigma^{33}=0.
\end{equation} 
By the continuity of stress tensor and assuming the thickness is small, we can assume $\sigma^{13}=\sigma^{23}=\sigma^{33}=0$ holds everywhere.
Making use of \eqref{planestress} and \eqref{energydensity}, the energy density in this situation is
\begin{equation}\label{2Denergydensity}
\epsilon=\frac{1}{2}\lambda \delta^{ij}\delta^{kl}u_{ij}u_{kl}+\mu \delta^{ij}\delta^{kl}u_{ik}u_{jl},
\end{equation}
where $i,j$ ranges from 1 to 2 and the effective 2D Lamé constants are
\begin{equation}\label{2DLame}
\lambda=\frac{2(B_{3D}-\frac{2}{3}\mu_{3D})\mu_{3D}}{B_{3D}+\frac{4}{3}\mu_{3D}},~\mu=\mu_{3D}.
\end{equation}
Thus, the effective 2D bulk modulus is
\begin{equation}\label{2DBulkmodulus}
B=\lambda+\mu=\frac{3B_{3D}\mu_{3D}}{B_{3D}+\frac{4}{3}\mu_{3D}}.
\end{equation}
Let $\vec{v}=v^1\vec{x}^1+v^2\vec{x}^2$ be the in-plane displacement, noting \eqref{2Denergydensity} has the same form as \eqref{energydensity}, the equation of motion for $\vec{v}$ is then
\begin{equation}\label{2Dctseom}
\rho\frac{\partial^2 \vec{v}}{\partial t^2}=(\lambda+\mu)\vec{\nabla}\vec{\nabla}\cdot\vec{v}+\mu \nabla^2\vec{v}.
\end{equation}
We simplify the equation again using the Helmholtz decomposition, giving
\begin{equation}\label{longandtranseom}
\begin{aligned}
&\rho\frac{\partial^2 \vec{v_t}}{\partial t^2}-\mu \nabla^2\vec{v_t}=0,\\
&\rho\frac{\partial^2 \vec{v_l}}{\partial t^2}-(2\mu+\lambda) \nabla^2\vec{v_l}=0.
\end{aligned}
\end{equation}
Now we calculate Rayleigh waves in the aforementioned thin sheet, but modified into a half-sheet laying in $x^2\leq 0$.
Let the longitudinal and transverse components of the Rayleigh wave be 
\begin{equation}\label{rayleighansatz}
\begin{aligned} 
&\vec{v_l}=e^{i(kx^1-\omega t)+\kappa_l x^2}(\alpha^1, \alpha^2)^T,\\
&\vec{v_t}=e^{i(kx^1-\omega t)+\kappa_t x^2}(\beta^1, \beta^2)^T.
\end{aligned}
\end{equation}
Plugging \eqref{rayleighansatz} into \eqref{longandtranseom}, we have
\begin{equation}
\kappa_l=\sqrt{k^2-\frac{\omega^2}{c_l^2}}, \; \kappa_t=\sqrt{k^2-\frac{\omega^2}{c_t^2}},
\end{equation}
where $c_l=\sqrt{\frac{\lambda+2\mu}{\rho}}$ and $c_t=\sqrt{\frac{\mu}{\rho}}$ are the speeds of longitudinal wave and transverse wave, respectively.
The $x^2=0$ line being an open boundary gives us the boundary conditions
\begin{equation}\label{ctsbdy}
\begin{aligned}
&\sigma^{12}=\mu(\frac{\partial v^1}{\partial x^2}+\frac{\partial v^2}{\partial x^1})=0,\\
&\sigma^{22}=\lambda\frac{\partial v^1}{\partial x^1}+(\lambda+2\mu)\frac{\partial v^2}{\partial x^2}=0.
\end{aligned}
\end{equation}
Plugging \eqref{rayleighansatz} into \eqref{ctsbdy}, we get
\begin{equation}
\begin{aligned}
& \kappa_l \alpha^1+ik\alpha^2+\kappa_t \beta^1+ik\beta^2=0,\\
&\frac{ik\lambda}{\lambda+2\mu}\alpha^1+\kappa_l \alpha^2+\frac{ik\lambda}{\lambda+2\mu}\beta^1+\kappa_t \beta^2=0.
\end{aligned}
\end{equation}
The curl free condition of $\vec{v_l}$ and divergence free condition of $\vec{v_t}$ give us
\begin{equation}
\begin{aligned}
-&\kappa_l \alpha^1+ik \alpha^2=0,\\
& ik\beta^1+\kappa_t \beta^2=0.
\end{aligned}
\end{equation}
We have 4 linear equations for 4 variables $\alpha^1, \alpha^2, \beta^1, \beta^2$. 
In order that the equations have non-zero solutions, the corresponding coefficient matrix needs to be of determinant zero, which gives us 
\begin{equation}
\xi^6-8\xi^4+(24-\frac{16 c_t^2}{c_l^2})\xi^2+\frac{16 c_t^2}{c_l^2}-16=0, \;\xi=\frac{\omega}{kc_t}.
\end{equation}
In terms of $h={B}/{\mu}$, $\xi$ satisfies
\begin{equation}
\xi^6-8\xi^4+(8+\frac{16h}{1+h})\xi^2-\frac{16h}{1+h}=0.
\end{equation}
The corresponding modes at the $x^2=0$ line are
\begin{equation}\label{ctsmodeshape}
\begin{aligned}
&\psi_{k}(x^1,0)=e^{i(kx^1-\omega t)}(\frac{1}{2}\xi\sqrt{1+\frac{1}{h}},\frac{1}{4}i\xi(2-\xi^2)\sqrt{\frac{1+h}{h-h\xi^2}})^T,\\
&\psi_{-k}(x^1,0)=e^{i(-kx^1-\omega t)}(\frac{1}{2}\xi\sqrt{1+\frac{1}{h}},-\frac{1}{4}i\xi(2-\xi^2)\sqrt{\frac{1+h}{h-h\xi^2}})^T.
\end{aligned}
\end{equation}

\section{Linear response to point source on the edge}
Here we derive the linear response theory for both 2D continuum and discrete systems (the same derivation applies to 3D systems) for systems subject to a point source of vibration on the edge.  

\subsection{Continuum version}
Consider the same homogeneous elastic thin sheet as described in Sec.~\ref{SEC:2DRayleigh}, we apply a local driving force $\vec{F}(r,t)=Re(\vec{F}e^{-i\omega t})\delta(r)$ to the boundary (the $(0, 0)$ point) of the system. 
The complex equation of motion is then
\begin{equation}\label{ctseom}
\begin{aligned}
&\rho\frac{\partial^2 \vec{v}}{\partial t^2}=(\lambda+\mu)\vec{\nabla}\vec{\nabla}\cdot\vec{v}+\mu \nabla^2\vec{v}+\vec{F}\delta(r)e^{-i\omega t},\\
&\vec{v}(r,0)=\frac{\partial \vec{v}}{\partial t}(r,0)=0.
\end{aligned}
\end{equation}
Let $\vec{\psi}_{k}(r)e^{-i\omega_k t}$ be Rayleigh waves with wave vector $k$ and $\vec{\phi}_{s}(r,t)$ be bulk waves with index $s$ (since the bulk waves are not important here we use an arbitrary index to identify them). 
The solution of \eqref{ctseom} can then be decomposed into
\begin{equation}\label{decomposition}
\vec{v}(r,t)=\int_{k}dk C_k(t)\vec{\psi}_{k}(r)e^{-i\omega_k t}+\int_{s}ds G_s(t)\vec{\phi}_{s}(r,t).
\end{equation}
Plugging \eqref{decomposition} into \eqref{ctseom}, applying $\vec{\psi}_{k}(r)e^{-i\omega_k t}$ to both sides, and doing an integration over time and position, we have
\begin{equation}\label{3dcomponent}
\begin{aligned}
\frac{d^2 C_{k}(t)}{dt^2}-&2i\omega_{j}(k)\frac{d C_{k}(t)}{dt}= \vec{\psi}_{k}(0)^*.\vec{F}\; e^{-i(\omega-\omega_{j}(k))t},\\
&\left.\frac{d C_{k}(t)}{dt}\right\vert_{t=0}=C_{k}(0)=0.
\end{aligned}
\end{equation}
For a given $k$, if $\omega\ne\omega_{k}$, $C_{k}(t)$ is bounded. 
However, if $\omega=\omega_{k}$, we have 
\begin{equation}
C_{j}(k,t)=\frac{i\vec{\psi}_{k}^*(0).\vec{F}}{\omega}t + q_{k}(t),
\end{equation}
where $q_{k}(t)$ is bounded. Therefore, after long times, most energy is in the two modes $\vec{\psi}_{k}(r)e^{-i\omega_k t}$ and $\vec{\psi}_{-k}(r)e^{-i\omega_k t}$. To maximally excite the $+k$ Rayleigh wave, the force applied to system needs to follow the trajectory of the motion corresponding to $\vec{\psi}_{k}(r)e^{-i\omega_k t}$, which means $\vec{\psi}_{k}(0)=\vec{F}$. 
In this case, the ratio $|{C_{-k}}/{C_{k}}|$, whose square quantifies the ratio between energy going in $+k$ direction and $-k$ direction, is then
\begin{equation}
\Big{\vert}\frac{C_{-k}}{C_{k}}\Big{\vert}=\Big{\vert}\frac{\vec{\psi}_{-k}(0)^*.\vec{\psi}_{k}(0)}{\vec{\psi}_{k}(0)^*.\vec{\psi}_{k}(0)}\Big{\vert}.
\end{equation}
Fig.~1(c) and Fig.~1(d) in the main text show $|{C_{-k}}/{C_{k}}|$ as a function of ${B_{3D}}/{\mu_{3D}}$.



\subsection{Discrete Version}
Consider a spring mass network that is periodic, and infinite in the $x^2$ direction and finite in the $x^1$ direction. 
For simplicity, we assume the lattice spacing in the $x^2$ direction to be $1$. 
Such system has two edge bands, and a finite number of bulk bands. 
We index the bands from low to high frequency by $1,2,3...2N$ (assuming N nodes per unit cell) and also index the nodes by $1,2,3...$ (The entire system has infinitely many nodes). 
Let $\vec{u}^m$ be the displacement vector of the $m^{th}$ node and $\vec{u}=(\vec{u}^1,...,\vec{u}^n...)$ be the displacements of the entire system. 
We also use $\vec{F}^i$ to denote the force on the $i^{th}$ node and $\vec{F}=(\vec{F}^1,...,\vec{F}^n...)$ be the forces on the entire system. 
We assume an external force $\vec{F}^i=(F^i_1\cos\omega t, F^i_2\cos(\omega t+\phi))$, where the frequency $\omega$ lies in the range of the edge bands, is applied to the $i^{th}$ node at the time $t=0$. 
Then, the equation of motion of the system together with initial conditions is
\begin{equation}
\label{realdiscreteeom}
m\vec{u}''=-D\vec{u}+\vec{F}, \; \vec{u}(0)=\vec{0}, \; \vec{u}'(0)=\vec{0}.
\end{equation}
To solve this equation, we rewrite $\vec{F}= Re(\vec{f}e^{-i\omega t}$) and solve the complex differential equation
\begin{equation}\label{complexdiscreteeom}
m\vec{u}''=-D\vec{u}+\vec{f}e^{-i\omega t}, \; \vec{u}(0)=\vec{0}, \; \vec{u}'(0)=\vec{0}.
\end{equation}
The real part of the solution of \eqref{complexdiscreteeom} is the solution of \eqref{realdiscreteeom}.
Let $\vec{\psi}_{j}(k)$ be the normalized Bloch state (where their inner product is the $\delta$ function) in the $j^{th}$ band with wave vector $k$ and corresponding eigenfrequency $\omega_{j}(k)$. 
The solution of the equation of motion can be expanded as
\begin{equation}\label{blochexpansion}
\vec{u}(t)=\int_{-\pi}^{\pi}C_{j}(k,t)e^{-i\omega_{j}(k)t}\vec{\psi}_{j}(k).
\end{equation}
Plugging this expansion into \eqref{realdiscreteeom}, we have
\begin{equation}
\begin{aligned}
\frac{d^2 C_{j}(k,t)}{dt^2}-&2i\omega_{j}(k)\frac{d C_{j}(k,t)}{dt}=\langle \vec{\psi}_{j}(k) | \vec{f} \rangle e^{-i(\omega-\omega_{j}(k))t}\\
&\left.\frac{d C_{j}(k,t)}{dt}\right\vert_{t=0}=C_{j}(k,0)=0.
\end{aligned}
\end{equation}
For a given $j,k$, if $\omega\ne\omega_{j}(k)$, $C_{j}(k,t)$ is bounded. 
However, if $\omega=\omega_{j}(k)$, we have 
\begin{equation}
C_{j}(k,t)=\frac{i\langle \vec{\psi}_{j}(k) | \vec{f} \rangle}{\omega}t + q_{j}(k,t),
\end{equation}
where $q_{j}(k,t)$ is bounded. 
Therefore, at long times, most energy is in the four modes $\vec{\psi}_{1}(k_1)$, $\vec{\psi}_{1}(-k_1)$, $\vec{\psi}_{2}(k_2)$,
and
$\vec{\psi}_{2}(-k_2)$. 
Without loss of generality, we can assume the $i^{th}$ node is on the left edge of the system and we want most energy localized at the edge propagating in $+x^2$ direction. 
The force applied to the $i^{th}$ node needs to follow the trajectory of the motion corresponding to some linear combination of modes $\vec{\psi}_{1}(k_1)$ and $\vec{\psi}_{2}(k_2)$ at node $i$. 
Mathematically, that means $\vec{f}$ maximizing 
\begin{equation}\label{maximazer}
\frac{|\langle \vec{\psi}_{1}(k_1) | \vec{f} \rangle|^2+|\langle \vec{\psi}_{2}(k_2) | \vec{f} \rangle|^2}{|\langle \vec{f}|\vec{f} \rangle|^2}=\frac{|\langle \vec{\psi}_{1}(k_1)^{i} | \vec{f}^{i} \rangle|^2+|\langle \vec{\psi}_{2}(k_2)^{i} | \vec{f}^{i} \rangle|^2}{|\langle \vec{f}^{i}|\vec{f}^{i} \rangle|^2}.
\end{equation}

\section{Orbital and spin angular momentum of phonons}

In this section, we discuss the general formulation of angular momentum of phonons as vector fields and their orbital and spin components. 
The main reference of this formulation is Ref.~\cite{francesco2012conformal}.

\subsection{Noether's theorem}
Let's start with a brief review of Noether's theorem in field theory, where each continuous symmetry leads to a conserved current.  

Consider a generic action $S$ of a set of fields $\phi_a$, where $a=1,\ldots,n$, such that
\begin{equation}
    S = \int dx \, \mathcal{L}\lbrack \phi_a , \partial_\mu \phi_a \rbrack,
\end{equation}
where we have kept in the Lagrangian to leading order derivatives of $\phi$.  
To make the notation compact, we use the 4D space-time coordinate $x^{\mu}\equiv (t,x^1,x^2,x^3)$, where the first component is time and the latter three are coordinates in 3D space.  
A generic infinitesimal transformation can be written as
\begin{align}\label{EQ:transform}
    x^{\mu} &\to  x'^{\mu}=x^{\mu}+\Delta x^{\mu}, \nonumber\\
    \phi_a(x) &\to \phi'_a(x') =\phi_a(x)+ \Delta \phi_a(x),
\end{align}
where $\Delta x$ and $\Delta\phi$ are both caused by the same transformation $w$, which we can generally write as
\begin{align}\label{EQ:Deltaw}
    \Delta x^{\mu} = \frac{\partial x^{\mu}}{\partial w_b} w_b, \nonumber\\
    \Delta \phi_a = \frac{\partial \phi_a}{\partial w_b} w_b .
\end{align}
Examples of such transformations can be found in the following subsections.  

The transformation of the spatial derivatives of the field follows
\begin{equation}
    \partial_\mu   \phi_a\to \partial_\mu '  \phi'_a = \partial_\mu   \phi_a - \partial_\nu   \phi_a \partial_\mu \Delta x^{\nu} +\partial_\mu \Delta\phi_a,
\end{equation}
where the second term on the right hand side is from the variation of $x$, and the last term is from the variation of $\phi$, as results of the transformation.  

At the same time, the integral over $dx'$ can be written as the integral over $dr$ with a Jacobian 
\begin{equation}
    dx\to dx'=dx \det (\partial_\mu  x'^{\nu})
    = dx (1+\partial_\mu \Delta x^{\mu}). 
\end{equation}

\begin{widetext}
Putting them all together, we have the transformed action
\begin{equation}
    S'=\int dx (1+\partial_\mu \Delta x^{\mu}) 
    \mathcal{L} \left\lbrack
    \phi_a(x)+ \Delta \phi_a(x) , 
    \partial_\mu   \phi_a - \partial_\nu   \phi_a \partial_\mu \Delta x^{\nu} +\partial_\mu \Delta\phi_a
    \right\rbrack.
\end{equation}
This action can be expanded to leading order in the transformation $\Delta x,\Delta \phi$, so that
\begin{equation}
    S'=\int dx 
     \left\lbrack \mathcal{L} + \partial_\mu \Delta x^{\mu} \mathcal{L} 
    +\frac{\partial \mathcal{L}}{\partial \phi_a} \Delta \phi_a(x) 
    - \frac{\partial \mathcal{L}}{\partial (\partial_\mu \phi_a)} \partial_\nu   \phi_a \partial_\mu \Delta x^{\nu}
    + \frac{\partial \mathcal{L}}{\partial (\partial_\mu \phi_a)} \partial_\mu \Delta \phi_a(x) 
    \right\rbrack.
\end{equation}
We can also use the Euler-Lagragian equation (the equation of motion of this field theory)
\begin{equation}
    \frac{\partial \mathcal{L}}{\partial \phi_a} 
    = \partial_\mu \left( \frac{\partial \mathcal{L}}{\partial (\partial_\mu \phi_a)} \right) 
\end{equation}
to simplify $S'$ so that 
\begin{equation}
    S'-S = \int dx 
     \left\lbrace  \left\lbrack 
     \delta^{\mu}_{\nu} \mathcal{L} 
     -\frac{\partial \mathcal{L}}{\partial (\partial_\mu \phi_a)} \partial_\nu   \phi_a 
     \right\rbrack \partial_\mu \Delta x^{\nu} 
    +\partial_{\mu}\left\lbrack \frac{\partial \mathcal{L}}{\partial (\partial_\mu \phi_a)} \Delta \phi_a(x) \right\rbrack
    \right\rbrace, 
\end{equation}
where the first bracket denotes the effect of (external) transformation of coordinate, and the second bracket denotes the effect of (internal) transformation of the field. 

Because both of these transformations are caused by the same source $w(x)$, we can use Eq.~\ref{EQ:Deltaw} to rewrite everything in terms of $w$.  
In particular, we have terms carrying $w$ and terms carrying $\partial_{\mu}w$.

Since Eq.~\ref{EQ:transform} can be viewed as simply a coordinate transformation (homogeneous or inhomogeneous), we must have $S'=S$ by definition.  
On the other hand, when the system has a continuous symmetry such as a homogeneous spatial translation or rotation, the Lagragian (not just the total action) must remain the same before and after the transformation.  
Thus the $w$ term must identically vanish everywhere.  
Therefore, to make $S'=S$, we must also have the integral of the $\partial_{\mu}w$ terms vanish, which are
\begin{equation}
    0=\int dx 
     \left\lbrace  \left\lbrack 
     \delta^{\mu}_{\nu} \mathcal{L} 
     -\frac{\partial \mathcal{L}}{\partial (\partial_\mu \phi_a)} \partial_\nu   \phi_a 
     \right\rbrack \frac{ \partial \Delta x^{\nu} }{\partial w_b}
    + \frac{\partial \mathcal{L}}{\partial (\partial_\mu \phi_a)} \frac{ \partial \Delta\phi_a }{\partial w_b} 
    \right\rbrace \partial_{\mu} w_b.
\end{equation}
Because $w$ can be an arbitrary transformation, we move the partial differential via integral by parts and find conserved currents
\begin{equation}\label{EQ:current}
    j^{\mu}_b = \left\lbrack 
     -\delta^{\mu}_{\nu} \mathcal{L} 
     +\frac{\partial \mathcal{L}}{\partial (\partial_\mu \phi_a)} \partial_\nu   \phi_a 
     \right\rbrack \frac{ \partial \Delta x^{\nu} }{\partial w_b}
    -\frac{\partial \mathcal{L}}{\partial (\partial_\mu \phi_a)} \frac{ \partial \Delta\phi_a }{\partial w_b},
\end{equation}
which satisfies $\partial_{\mu}j^\mu_b=0$. 

Now we have proved Noether's theorem, where each continuous symmetry $w_b$ leads to a conserved current $j^\mu_b$ satisfying $\partial_{\mu}j^\mu_b=0$.  
Next we discuss its consequence on spatial translations and rotations, the cases of interest in this paper.
\end{widetext}

\subsection{Translational symmetry}
The case of spatial translations can be simply written as translations in the three spatial directions $w=(w_1,w_2,w_3)$.
Thus $\frac{ \partial \Delta x^{\nu} }{\partial w_b} =0$ for $\nu=0$ (time), and $\delta_{b}^{\nu}$ for $\nu=1,2,3$ (space).  
The $\frac{ \partial \phi_a }{\partial w_b}$ terms all vanish as the field itself doesn't transform under spatial translation.  

Plugging these into the formula of conserved current [Eq.~\eqref{EQ:current}], we find
\begin{align}
    j_b^0 &= \frac{\partial \mathcal{L}}{\partial (\partial_{0} \phi_a)} \partial_{b} \phi_a , \\
    j_b^i &= -\delta_b^i \mathcal{L} +\frac{\partial \mathcal{L}}{\partial (\partial_{i} \phi_a)} \partial_{b} \phi_a,
\end{align}
where index $i=1,2,3$ runs through spatial indices.  

Now we restore the notation where time and space are separated (instead of the 4D notation), and since all indices are in Euclidean space, we will have them all as sub-indices.  
We have the resulting continuity equation
\begin{equation}\label{EQ:MomentumC}
    \partial _t g_m + \partial _n \tau_{mn} =0,
\end{equation}
which describes \emph{momentum conservation} of this field theory.  
Here the momentum density $g$ and momentum flow $\tau$ are given by
\begin{align}
    g_m \equiv -j_m^0 , \\
    \tau_{mn} \equiv -j_m^n.
\end{align}
This continuity equation [Eq.~\eqref{EQ:MomentumC}] tells us that as a result of spatial translational symmetry, the total momentum of the system
\begin{align}
    P_m = \int d^3 x  g_m (x)
\end{align}
is a conserved quantity.

\subsection{Rotational symmetry}

The case of spatial rotations, which is of central interest to our discussion here, is slightly more complicated. 
An infinitesimal spatial rotation can be written as $x \to x' = (E+I \cdot w)  x$, where $E$ is identity matrix, $I$ denotes the 3 generators of $SO(3)$, and $w=(w_1,w_2,w_3)$ are the angles of rotation in 3D along $x,y,z$, respectively.  
This can be written in a more explicit form
\begin{equation}
    x'_i = x_i + \varepsilon_{ib j} \, r_j \, w_b, 
\end{equation}
where $\varepsilon$ is the Levi-Civita symbol.  
Thus, we have $\frac{\partial \Delta x_i}{\partial w_{b}} = \varepsilon_{ib j}  \, x_j$ for the spatial degrees of freedom and $\Delta x^0=0$.  

How does the field $\phi$ transform under spatial rotation?
This depends on the type of field we consider.  
Scalar fields won't change.  
Vector fields, the case of interest here as elasticity is concerned with displacements of components in a solid, should rotate as a vector.  
This tells us that $\frac{\partial \Delta\phi_a}{\partial w_{b}}= \varepsilon_{abc}  \, \phi_c$.  

Plugging these into the formula of conserved current [Eq.~\eqref{EQ:current}], we find that for the case of rotation
\begin{align}
    j_b^0 &= \frac{\partial \mathcal{L}}{\partial (\partial_{0} \phi_a)} \partial_{\nu} \phi_a \varepsilon_{\nu b m}  \, x_m 
    - \frac{\partial \mathcal{L}}{\partial (\partial_{0} \phi_a)} 
    \varepsilon_{a b m}  \, \phi_m, \\
    j_b^i &= \left\lbrack -\delta_{\nu}^i \mathcal{L} +\frac{\partial \mathcal{L}}{\partial (\partial_{i} \phi_a)} \partial_{\nu} \phi_a \right\rbrack \varepsilon_{\nu b m}  \, x_m
    - \frac{\partial \mathcal{L}}{\partial (\partial_{i} \phi_a)} 
    \varepsilon_{a b m}  \, \phi_m,
\end{align}
where index $i=1,2,3$ runs through spatial indices (and we again wrote them all as sub-indices for simplicity).  
From the continuity equation, $\partial_{\mu}j^{\mu}=0$, we know that $j_b^0$ offers a new conserved quantity.  
In particular, this can be written into two parts: the first term, which is from the rotation of the coordinate, and the second term which is from the rotation of the field,
\begin{align}
    - j_b^0 &= l_b + s_b,
\end{align}
where 
\begin{align}
    l_b &\equiv \varepsilon_{bm\nu} x_m g_{\nu} ,
    \\
    s_b &\equiv \varepsilon_{bma} \phi_m \frac{\partial \mathcal{L}}{\partial (\partial_{t} \phi_a)} .
\end{align}
These two quantities are the orbital and spin angular momenta of the vector field $\phi$.  
This can also be written as 
\begin{align}
    \vec{l} &\equiv  \vec{x} \times \vec{g}, \label{EQ:orbital}
    \\
    \vec{s} &\equiv  \vec{\phi} \times \frac{\partial \mathcal{L}}{\partial (\partial_{t} \vec{\phi})} .
\end{align}

\subsection{Angular momentum of elastic waves}

Now we apply this formalism to elasticity, to derive the expressions for the orbital and spin angular momenta.  

The Lagrangian of a general isotropic solid can be written as
\begin{equation}
    \mathcal{L} = \frac{\rho}{2} \partial_t u_i \partial_t u_i
    - \frac{\lambda}{2} (\partial_i u_i )^2 - \frac{\mu}{2}(\partial_i u_j \partial_i u_j +\partial_i u_j \partial_j u_i)^2,
\end{equation}
where $\rho$ is the mass density, $\mu,\lambda$ are the Lame coefficients.  
Identifying $\vec{u}$ as the vector field $\vec{\phi}$ discussed above, this leads to densities of momentum $g$, orbital angular momentum $l$ and spin angular momentum $s$, where
\begin{align}
    g_i &= -\rho \partial_t u_j \partial_i u_j , \\
    l_i &= \varepsilon_{ijk} x_j g_k , \\
    s_i &= \rho \varepsilon_{ijk} u_j  \partial_t u_k .
\end{align}
In the vectorial form, the spin angular momentum can also be written as
\begin{align}\label{EQ:spin}
    \vec{s} &= \rho \vec{u}\times (\partial_t \vec{u}).
\end{align}
From this formula it is clear that the spin angular momentum of these elastic waves characterize the angular momentum of mass points rotating around their mean positions.  
In contrast the orbital angular momentum comes from the wave momentum $\vec{k}$.  

We can apply them to a generic plane-wave field
\begin{equation}
    \vec{u}(\vec{x},t) = \Re\left(\vec{A} \, e^{i(\vec{k}\cdot \vec{x}-\omega t)} \right),
\end{equation}
where the amplitude $A$ can be a complex vector denoting possible phase difference of different wave components ($u_x,u_y,u_z$).
This leads to
\begin{align}
    \vec{g} (\vec{x},t) &= \rho \omega   
    \left\vert \Im \left(\vec{A} e^{i(\vec{k}\cdot \vec{x}-\omega t)}\right)\right\vert^2 \vec{k}\\
    \vec{l} (\vec{x},t) &=\rho \omega    \left\vert \Im \left(\vec{A} e^{i(\vec{k}\cdot \vec{x}-\omega t)}\right)\right\vert^2 \vec{x} \times \vec{k},  \\
    \vec{s} (\vec{x},t) &=\frac{1}{2}\rho \omega \Im (\vec{\overline{A}}\times \vec{A}),
\end{align}
where $\Re$ and $\Im$ denote the real and imaginary parts, respectively, and $\overline{A}$ is the complex conjugate of $A$. 
From this form, it is clear that $\vec{\overline{A}}\times \vec{A}$ has to be complex in order for this wave to carry a spin angular momentum.

\subsection{Spin angular momentum of elastic waves in 2D and 3D}

We first use these formula to examine the spin of elastic waves in the bulk of isotropic materials.  
In 3D, it is obvious that longitudinal waves don't carry spin: their $\vec{A}$ is along the wave vector $\vec{k}$ so there is only one direction of vibration, leading to $\vec{\overline{A}}\times \vec{A}=0$.  

Transverse waves in 3D isotropic solids can be generally written as $\vec{A}=\{A_1,A_2,0\}$, taking the direction of $k$ to be the third dimension.  
Without losing generality we can normalize the amplitude to be unity and factorize out a common phase, so that $\vec{A}=\{1, e^{i\theta},0\}$.  
Therefore $\vec{\overline{A}}\times \vec{A}=\{0,0,\sin\theta\}$, which is along $\vec{k}$ with an amplitude between $[-1,1]$.  
Thus $\vec{s} = \frac{1}{2}\rho\omega\sin\theta e_{\vec{k}}$ where $e_{\vec{k}}$ is the unit vector along $\vec{k}$. 
It is straightforward to see that circularly polarized transverse waves maximize $\vert \vec{s}\vert$.  

In 2D, there are only 2 directions for vibrations, and the wave vector $\vec{k}$ is in the same plane.  
The wave equation can be generally written as ~\eqref{longandtranseom},
where it is clear that, for materials with finite bulk modulus $B>0$ and shear modulus $\mu>0$, the longitudinal and transverse waves have different dispersion relations and can't combine to carry any angular momenta in the bulk.  

In the special case of $B=0$ in 2D, the two waves have the same wavespeeds (and linear dispersion relations). 
They can be written in a similar form as the 3D case, leading to 
\begin{equation}
    \vec{s} = \frac{1}{2}\rho\omega\sin\theta e_{\vec{z}},
\end{equation}
describing spin angular momentum along the normal direction of the 2D plane.  

When quantizing this problem, the spin angular momentum of phonons as quanta of these waves take values of $(-\hbar,0,+\hbar)$.  
In 3D, this is similar to spins of photons, but the definition of $S_z$ in 2D is unique to $B=0$ materials. 

Interestingly, near the edge of a 2D sheet, Rayleigh waves combine both longitudinal and transverse components, and carry spin angular momentum, as we discuss in the main text.  It also becomes clear here that the same wavespeeds of longitudinal and transverse waves in the $B=0$ case allows the spin of the Rayleigh wave extend into the bulk (Fig.~1 in the main text).





\section{Backscattering-free edge waves and winding number in 2D $B/\mu\ll 1$ continuum}


In this section, we discuss in more detail the complex analytic function formulation we used to describe backscattering-free edge waves in 2D $B/\mu\ll 1$ continuum, and the definition of the new winding number.

Consider a 2D homogeneous elastic material with effective bulk modulus $B=0$ and effective Lamé constants $\lambda=-\mu$ in a bounded simply connected region $\Omega\subset \mathbb{C}$. 
The equation of motion together with stress free boundary conditions under $B=0$ is
\begin{equation}
\begin{aligned}
&\rho\frac{\partial^2 \vec{v}}{\partial t^2}=\mu\nabla^2\vec{v},\\
&n^1\sigma^{11}+n^2\sigma^{12}=0,\\
&n^1\sigma^{12}+n^2\sigma^{22}=0,
\end{aligned}
\end{equation}
where $\vec{n}=(n^1,n^2)$ is the vector normal to the boundary (pointing outward).
The in plane displacement vector field $\vec{v}=v^1\vec{x}_1+v^2\vec{x}_2$ can be identified with a complex scalar field $\psi$ via $\psi=v^1+iv^2$. 
We can rewrite the boundary condition as
\begin{equation}
n^1\sigma^{11}+n^2\sigma^{12}+i(
n^1\sigma^{12}+n^2\sigma^{22})=0.
\end{equation}
Under the identification above, the equation of motion together with stress free boundary conditions under $B=0$ is
\begin{equation}\label{2dcomplexeom}
\begin{aligned}
\rho\frac{\partial^2\psi(z,t)}{\partial t^2}&=4\mu \frac{\partial^2\psi(z,t)}{\partial z \partial z^*}, z\in \Omega,\\
\frac{\partial \psi(z,t)}{\partial z^*}&=0, z\in \partial\Omega,
\end{aligned}
\end{equation}
where $\frac{\partial}{\partial z}=\frac{1}{2}(\frac{\partial}{\partial x}-i\frac{\partial}{\partial y})$, $\frac{\partial}{\partial z^*}=\frac{1}{2}(\frac{\partial}{\partial x}+i\frac{\partial}{\partial y})$, and $\partial \Omega$ is the boundary of $\Omega$.
Note that \eqref{2dcomplexeom} admits separation of variables
\begin{equation}\label{spinupsol}
\psi(z,t)=\phi(z)e^{i\omega t}
\end{equation}
and
\begin{equation}\label{spindownsol}
\psi(z,t)=\phi(z)e^{-i\omega t},
\end{equation}
where \eqref{spinupsol} corresponds to spin-up states (rotate counterclockwise) and \eqref{spindownsol} corresponds to spin-down states (rotate clockwise). 
The profile $\phi(z)$ satisfies
\begin{equation}\label{profileeq}
\begin{aligned}
-\omega^2\rho\phi(z)&=4\mu \frac{\partial^2\phi}{\partial z \partial z^*}(z), z\in \Omega,\\
\frac{\partial \phi}{\partial z^*}(z)&=0, z\in \partial\Omega.
\end{aligned}
\end{equation}
To solve \eqref{profileeq}, we let 
\begin{equation}\label{defineh}
h(z)=\frac{\partial \phi}{\partial z^*}.
\end{equation}
By taking $\frac{\partial}{\partial z^*}$ on both sides of \eqref{profileeq}, we see $h(z)$ satisfies Helmholtz equation with Dirichlet boundary conditions~\cite{strauss2007partial}
\begin{equation}\label{eomofh}
\begin{aligned}
-\frac{\omega^2\rho}{4\mu} h(z)&=\frac{\partial^2 h}{\partial z\partial z^*}(z), z\in \Omega, \\
h(z)&=0, z\in \partial\Omega.
\end{aligned}
\end{equation}
Since $\Omega$ is a bounded region, it is follows that the eigenvalues $-\frac{\omega^2\rho}{4\mu}$ are discrete. 
When $\omega\ne 0$, $h(z)$ has large variation in the bulk and $\phi(z)=-\frac{4\mu}{\omega^2\rho}\frac{\partial h}{\partial z}$ has large value throughout the bulk, hence it is not an edge mode.
When $\omega=0$, $h(z)$ has unique solution $h(z)\equiv 0$, hence $\frac{\partial \phi}{\partial z^*}(z)=0$, which is equivalent to $\phi(z)$ being a bounded analytic function.
Here, bounded means
\begin{equation}
\exists R>0, |\phi(z)|<R, \forall z\in\Omega
\end{equation}
and analytic means the limit
\begin{equation}
\lim_{u\rightarrow 0}\frac{\phi(z+u)-\phi(z)}{u}
\end{equation}
exists for all $z\in\Omega$.
For $B>0$, the equation of motion together with the boundary condition becomes
\begin{equation}\label{2dcomplexeomgeneral}
\begin{aligned}
\frac{\rho}{\mu}\frac{\partial^2\psi(z,t)}{\partial t^2}&=\frac{B}{\mu}(2\frac{\partial^2 \psi(z,t)}{\partial z\partial z^*}+\frac{\partial^2 \psi^*(z,t)}{\partial z^2})+4 \frac{\partial^2\psi(z,t)}{\partial z \partial z^*}, z\in \Omega,\\
\frac{\partial \psi(z,t)}{\partial z^*}&=\frac{B}{2\mu}\frac{n^1-in^2}{n^1+in^2}(\frac{\partial \psi(z,t)}{\partial z}+\frac{\partial \psi^*(z,t)}{\partial z^*}), z\in \partial\Omega.
\end{aligned}
\end{equation}
In the $B\ll\mu$ limit, the $B/\mu$ term can be viewed as a perturbation. 
Let $\psi_{\omega(B/\mu)}(z,t)$ be an eigenmode with frequency $\omega(B/\mu)$, where
\begin{equation} \lim_{B/\mu\rightarrow 0}\omega(B/\mu)=0.
\end{equation}
We have 
\begin{equation}
\lim_{B/\mu\rightarrow 0}\psi_{\omega(B/\mu)}(z,t)=\phi(z)
\end{equation}
for some bounded analytic function $\phi(z)$. If we start from $B/\mu=0$, then allow some small $B/\mu$, since  $\phi(z)e^{i\omega(B/\mu) t}$ and $\phi(z)e^{-i\omega(B/\mu) t}$ serve as zeroth order (in $B/\mu$) solution to \eqref{2dcomplexeomgeneral}, we see 
\begin{equation}
\psi_{\omega(B/\mu)}(z,t)\approx\phi(z)e^{i\omega(B/\mu) t}
\end{equation}
or we have 
\begin{equation}
\psi_{\omega(B/\mu)}(z,t)\approx\phi(z)e^{-i\omega(B/\mu) t}.
\end{equation}
Now we prove the following claim: for modes $\phi(z)$ giving rise to waves $\psi_{\omega(B/\mu)}(z,t)$ localized at the boundary, we can find $z_0\in\Omega$ such that $\phi(z_0)=0$ and $\phi(z)\ne 0$ when $z\in \partial\Omega$.  
The claim is a consequence of Rouché's theorem~\cite{gamelin2003complex}, which is: 
\begin{thm}
Let $\Omega$ be a bounded domain with piecewise smooth boundary $\partial\Omega$. 
Let $\phi(z)$ and $g(z)$ be two analytic functions that can be analytically extended to an open set containing $\Omega\cup\partial\Omega$. 
If $|g(z)|<|\phi(z)|, \forall z\in\partial\Omega$, then $\phi(z)$ and $\phi(z)+g(z)$ have the same number of zeros in $\Omega$, counting multiplicities.
\end{thm}
Mode localized at the boundary has small displacement in the bulk and large displacement at all boundaries, and hence we can find $z_1\in\Omega$ such that 
\begin{equation}\label{rouchecondition}
0<|\phi(z_1)|<|\phi(z)|, \forall z\in\partial\Omega.
\end{equation}
We then define $g(z)\equiv-\phi(z_1)$ and consider function
\begin{equation}
f(z)=\phi(z)+g(z).
\end{equation}
Since $\phi(z)$ and $g(z)$ satisfy \eqref{rouchecondition}, we can apply Rouché's theorem and see if $f(z)$ and $\phi(z)$ have the same number of zeros (counting multiplicity) in $\Omega$. 
Since $z=z_1$ is a zero of $f(z)$ in $\Omega$, we know $\phi(z)$ has at least one zero in $\Omega$.\\

Now we look at the behavior of profile $\phi(z)$ at the boundary. 
By argument principle~\cite{gamelin2003complex}, 
\begin{equation}\label{argumentprinciple}
\frac{1}{2\pi i}\int_{\partial\Omega}d\ln(\phi(z))=\frac{1}{2\pi i}\int_{\partial\Omega}\frac{\phi'(z)}{\phi(z)}dz=N_0-N_{\infty},
\end{equation}
where $\frac{1}{2\pi i}\int_{\partial\Omega}d\ln(\phi(z))$ is the number of counterclockwise turns $\phi(z)$ rotates when walking counterclockwise along the boundary (this means $\Omega$ is always on the left hand side), $N_0$ is the number of zeros of $\phi(z)$ (counting multiplicity) inside $\Omega$, and $N_{\infty}$ is the number of poles of $\phi(z)$ (counting multiplicity) inside $\Omega$. 
We note that $\phi(z)$ being bounded in $\Omega$ implies $N_{\infty}=0$ and $\phi(z_0)=0$ implies $N_0\geq 1$. 
Hence when walking counterclockwise along the boundary, $\phi(z)$ rotates counterclockwise on average. 
From this, we can see the phase zero point on the boundary in the spin-up solution $\phi(z)e^{i\omega(B/\mu) t}$ moves clockwise and the phase zero point on the boundary in the spin-down solution $\phi(z)e^{-i\omega(B/\mu) t}$ moves counterclockwise, spin and momentum are coupled in $0<B\ll\mu$ material. 
Note that this argument doesn't depend on the shape of the material, implying the wave will not be backscattered by defects on the boundary.

\section{Mode analysis in 2D $B/\mu\ll 1$ continuum materials}

Here we derive qualitative behaviors of the eigenfrequencies and eigenmode shapes for a homogeneous 2D $B/\mu\ll 1$ continuum material, with Lamé constants $\lambda=-\mu$, within an arbitrary shaped, bounded simply connected region $\Omega\subset\mathbb{C}$.

First, we examine its high frequency behavior. 
When $B/\mu=0$, from \eqref{eomofh} we see $\rho\omega^2/\mu$ is the eigenvalue of the laplacian operator $-\nabla^2$ with Dirichlet boundary conditions. Weyl's law states that
\begin{equation}\label{welylaw}
\lim_{n\rightarrow\infty}\frac{N(E)}{E}=\frac{|\Omega|}{2\pi},
\end{equation}
where $N(E)$ is the number of eigenvalues below $E$ and $|\Omega|$ is the area of $\Omega$. 
Let $\omega_n$ be the $n^{th}$ eigenfrequency, it follows from Weyl's law~\cite{strauss2007partial} that
\begin{equation}\label{highfrequencytrend}
\omega_n\approx \sqrt{\frac{2\pi n\mu}{|\Omega|\rho}}\propto \sqrt{n},
\end{equation}
when $n$ is sufficiently large. 
When $0<B/\mu\ll 1$, $\omega_n\propto\sqrt{n}$ should still hold as $0<B/\mu\ll 1$ is only a small perturbation.  When the nonzero eigenvalue $\rho\omega^2/\mu$ is present, the eigenfunction $h(z)$ exhibits a large value in the bulk. From \eqref{defineh}, we can deduce that the mode shape $\phi(z)$ also exhibits a large value in the bulk, hence confirming it as a bulk mode. 
Let $E_1$ be the smallest eigenvalue of \eqref{eomofh}, in nonzero but $B/\mu\ll 1$ limit, the bulk mode first appears at frequency $\sqrt{\frac{E_1\mu}{\rho}}$.

Second, we exam frequency behavior below $\sqrt{\frac{E_1\mu}{\rho}}$ for $0<B/\mu\ll 1$. 
Let $k_n$ be the wave vector of the $n^{th}$ mode along the boundary, we then have
\begin{equation}
k_n L=2\pi N_0(n),
\end{equation}
where $L$ is the perimeter of $\Omega$ and $N_0(n)$ is the number of turns $\phi_n(z)$ (the $n^{th}$ mode) winds.  
From
\begin{equation}
\frac{\omega_n}{k_n}=v_{phase},
\end{equation}
where 
\begin{equation}
\begin{aligned}
&v_{phase}=\xi\left(\frac{B}{\mu}\right)\sqrt{\mu/\rho}, 
\end{aligned}    
\end{equation}
is the phase velocity satisfying
\begin{equation}
\begin{aligned}
&\xi\left(\frac{B}{\mu}\right)^6-8\xi\left(\frac{B}{\mu}\right)^4+(8+\frac{16\frac{B}{\mu}}{1+\frac{B}{\mu}})\xi\left(\frac{B}{\mu}\right)^2-\frac{16\frac{B}{\mu}}{1+\frac{B}{\mu}}=0,
\end{aligned}    
\end{equation}
for Rayleigh waves, 
we know 
\begin{equation}
\omega_n=\frac{2\pi N_0(n) v_{phase}}{L}.
\end{equation}
Since the shape doesn't have any spatial symmetry, $\omega_n$, in general, can only have double degeneracy (corresponding to the degeneracy of spin up and spin down states) and the $(2n-1)^{th}$ edge mode and the $2n^{th}$ edge mode share a common mode shape $\phi_n(z)$. 
Assuming there is only one $\phi_n(z)$ for each $n$ (we have not been able to establish a rigorous relation for this), we then have $N_0(n)=\frac{n-2}{2}$ for even $n>3$, $N_0(n)=\frac{n-3}{2}$ for odd $n>3$ and $N_{0}(n)\approx\frac{n}{2}$. 
Therefore, at frequency below $\sqrt{\frac{E_1\mu}{\rho}}$, we have 
\begin{equation}
\omega_n=\frac{\pi nv_{phase}}{L}\propto n.
\end{equation}
The simulation in Fig.~2(a) of the main text uses $L=1.48m$, $B_{3D}/\mu_{3D}=0.01$. 
In that case, $v_{phase}=21.33$ m/s and
\begin{equation}
    \omega_n=\frac{\pi n v_{phase}}{L}=45.36n,
\end{equation}
which agrees with the simulated fitting slope ($43.33$) of the green dots in Fig.~3(a).
Since we have $N_0(n)\approx\frac{n}{2}$, the mode shape of the $(2n-1)^{th}$ and $2n^{th}$ edge mode $\phi_n(z)$ has $n$ zeros according to \eqref{argumentprinciple}. 
Hence,
\begin{equation}
\phi_n(z)=a_nz^n+...+a_1z+a_0
\end{equation}
with some large $a_n$, because the $n$ zeros lie inside $\Omega$. 
Since the function $z^n$ is more localized at $\partial\Omega$ than $z^m$ for $m<n$, we know that high frequency edge modes are more localized, which agrees with Fig.~3(b) in the main text.

\begin{widetext}
\section{Long wavelength behavior in Maxwell lattices}

\begin{figure*}[ht]
    \includegraphics[width=0.7\linewidth]{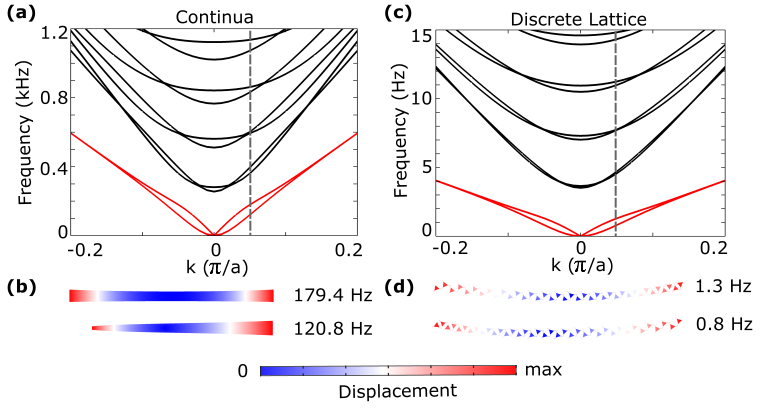}
    \caption{Dispersion relations of a 2D material strip in the low-$k$ regime for {\bf(a)} a continua strip  ($B_{3D}/\mu_{3D}=0.1$) and {\bf (c)} a discrete lattice ($B_{3D}/\mu_{3D}=0.173$), where the strip is free on left and right boundaries and periodic on the top and bottom. {\bf (b,d)} Mode profiles at $k = 0.05 {\pi}/{a}$.
    }
    \label{FigS_dispersion}
\end{figure*}

In Fig.~4(b) in the main text, we have identified two seemingly overlapping branches, which we denoted as Rayleigh modes. We note that, particularly at long wavelengths (low $k$), they can also be considered as Lamb modes due to the presence of the two free surfaces (right and left sides of the simulated domain shown in main text Fig.~4(d)). In Fig.~\ref{FigS_dispersion}(a,c), we show the dispersion for two 2D strips of moduli $B_{3D}/\mu_{3D}>0$ with periodic boundary conditions on the top and bottom and free boundary conditions on the left and right surfaces (simulated geometries shown in Fig.~\ref{FigS_dispersion}(b,d)). Figures~\ref{FigS_dispersion}(a,b) are of a continuum and Figs.~\ref{FigS_dispersion}(c,d) are of the same Maxwell lattice strip considered in the main text Fig.~4(b-d), but focusing in on low frequencies and wavenumbers. At low $k$, we can now see that the two Rayleigh modes split into two Lamb modes: one characterized by antisymmetric behavior and parabolic dispersion, and another characterized by symmetric behavior with linear dispersion. At low $k$, the symmetric mode approaches the longitudinal wavespeed. 
The higher branches, which we referred to as bulk modes, are also commonly known as ``higher order Lamb modes.'' 

There are a few noteworthy properties concerning these Lamb waves at small $k$.
First, a significant distinction exists between 2D and 3D scenarios. In 3D, Lamb waves always exhibit quadratic dispersion at small $k$, regardless of the value of $B$. 
However, in 2D, Lamb waves only arise when $B$ is finite. At $B=0$ in 2D, there is no hybridization between the Rayleigh waves at the two edges, and thus Lamb waves are not observed. Second, the issue of hybridization between modes localized to two independent edges or domain walls is not unique to our study or system. 
Such hybridization arises in various systems, including quantum Hall and quantum valley Hall edge states. When two 1D edges or domain walls are in close proximity, their hybridization leads to the formation of symmetric and anti-symmetric modes, subsequently resulting in a loss of one-way transport. Often such effects are neglected, parameters and geometries are chosen where this hybridization is weak. In our study, we follow the same approach, ensuring the parameter range studied avoids significant hybridization, namely by avoiding small $k$ and approaching $B=0$. However, it is important to note that except for the continuum case idealized to $B=0$ or infinitely small wavelengths, such hybridization will contribute to reduction of the edge mode backscattering immunity.
\end{widetext}

\section{Additional experimental details}

\subsection{Mortise and tenon assembly}

\begin{figure}[h]
    \centering
    \includegraphics[width=0.9\linewidth]{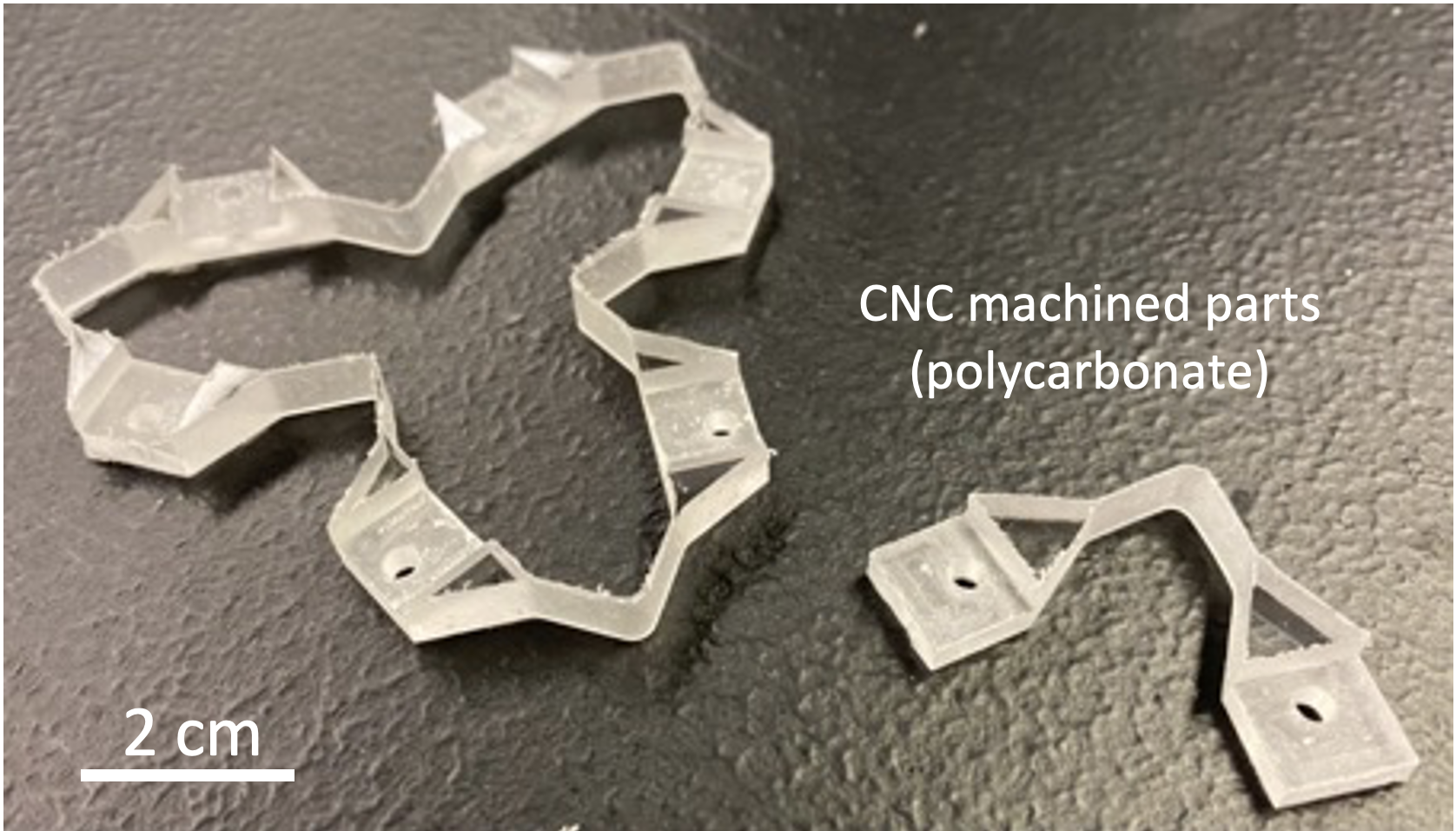}
    \caption{
    Example of CNC machined sublattice elements with mortise and tenon structures to form solid triangles.
    }
    \label{FigS_sublattice_elements}
\end{figure}

An example of CNC machined polycarbonate parts for lattice assembly via press fit is shown in Fig.\ref{FigS_sublattice_elements}.

\subsection{Energy ratio calculation}

\begin{figure}[h]
    \centering
    \includegraphics[width=0.9\linewidth]{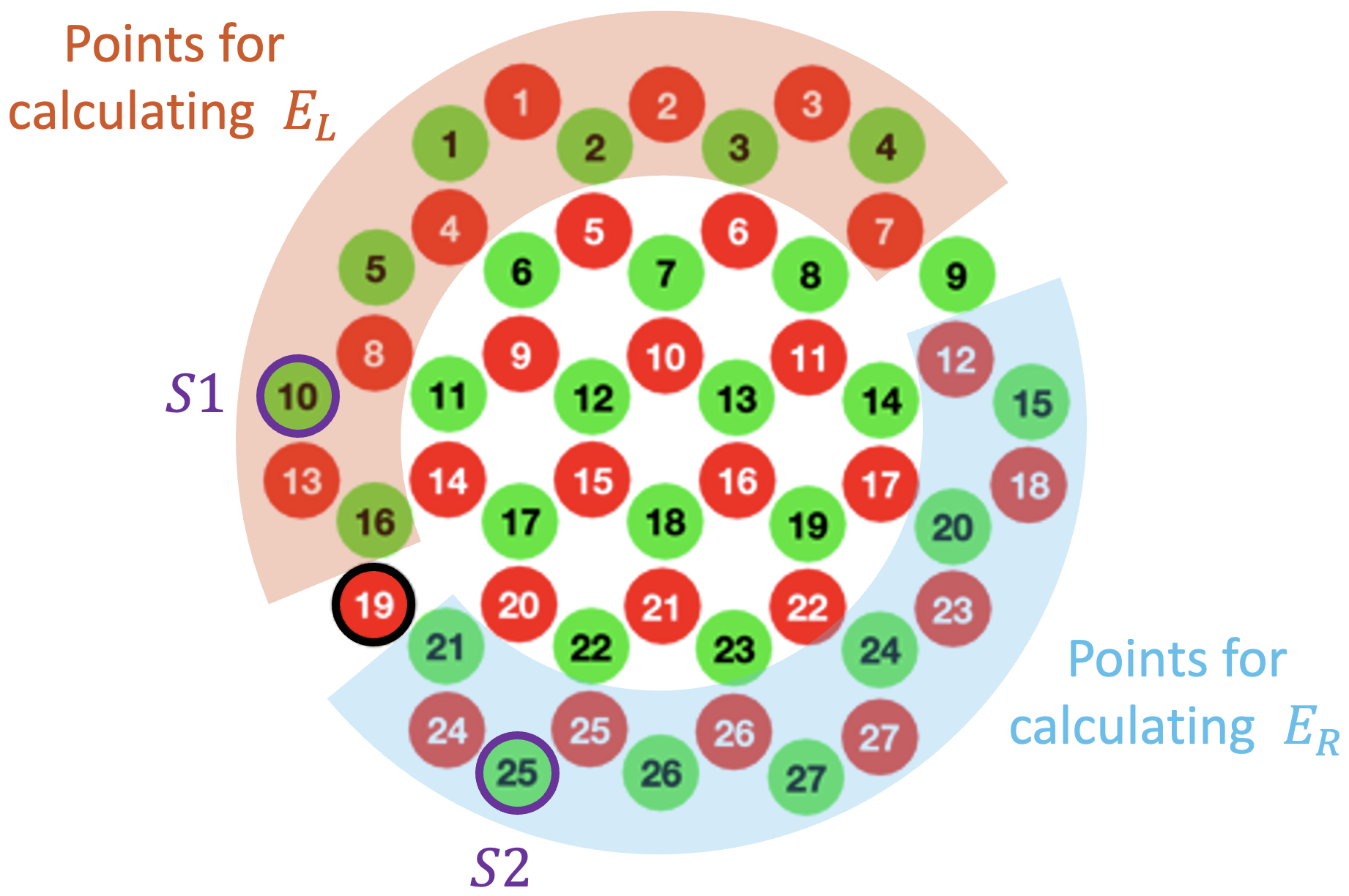}
    \caption{Triangles on the edge to the left and right of the excitation point (black circle) used for calculating energy ratio $E_L/E_R$ shown in Fig.~4(e-j) and plotting average kinetic energy ($v^2$) in Fig.~\ref{FigS_Energy ratio}. Numbers in the red and green solid circles are indices used for data processing in MATLAB. $S1$ and $S2$ mark the triangles on the edges equidistant from the actuated point used for tracking the trajectory phases shown in Fig.~\ref{FigS_Energy ratio}. 
    }
    \label{FigS_points_for_calculating_energy_ratio}
\end{figure}

The energy ratio calculated in experiments between the edge waves propagating in opposite directions is calculated by comparing the sums of the squared velocities of all edge triangles' centroids on the left and right sides of the actuator (shown in Fig.~\ref{FigS_points_for_calculating_energy_ratio}) over the time period of $0.01-0.15$ s, before the two edge waves encounter each other. It should be noted that this ratio does not correspond to the ratio $|C_{-k}/C_{k}|$, which predicts the relative amplitude of the opposite propagating modes in the $t \to \infty$. 

\subsection{Poisson's ratio measurement}

\begin{figure}[h]
    \centering
    \includegraphics[width=0.9\linewidth]{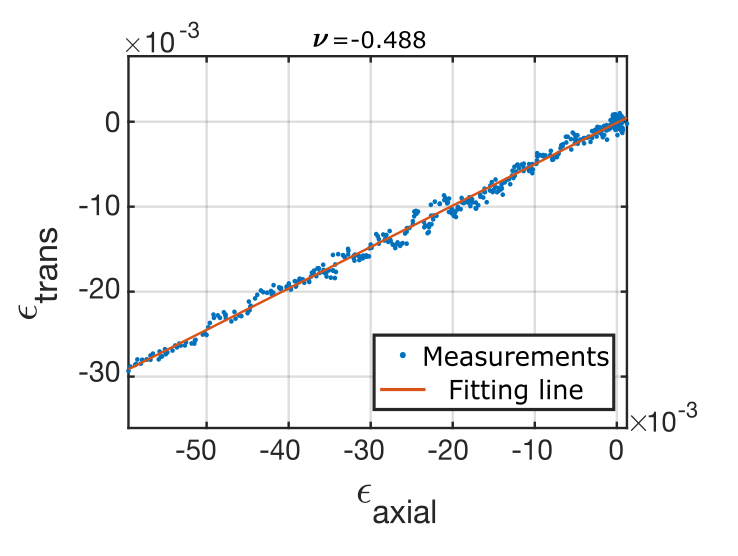}
    \caption{Measured Poisson's ratio of the auxetic Maxwell lattice. $\epsilon_{axial}$ and $\epsilon_{trans}$ represent the strains along and perpendicular to the compression, respectively.
    }
    \label{FigS_Poisson's ratio}
\end{figure}

The Poisson's ratio of our manufactured Maxwell lattice was measured based on manual compression of the lattice and image tracking of axial compression (perpendicular to the edge) and transverse (lateral) expansion.
The fitted slope gives $\nu\simeq-0.488$ (Fig.~\ref{FigS_Poisson's ratio}), corresponding to $B/\mu=0.344$ ($B_{3D}/\mu_{3D}=0.173$).

\subsection{Additional experiment and simulation comparison}

\begin{figure*}[ht]
    \centering
    \includegraphics[width=1\linewidth]{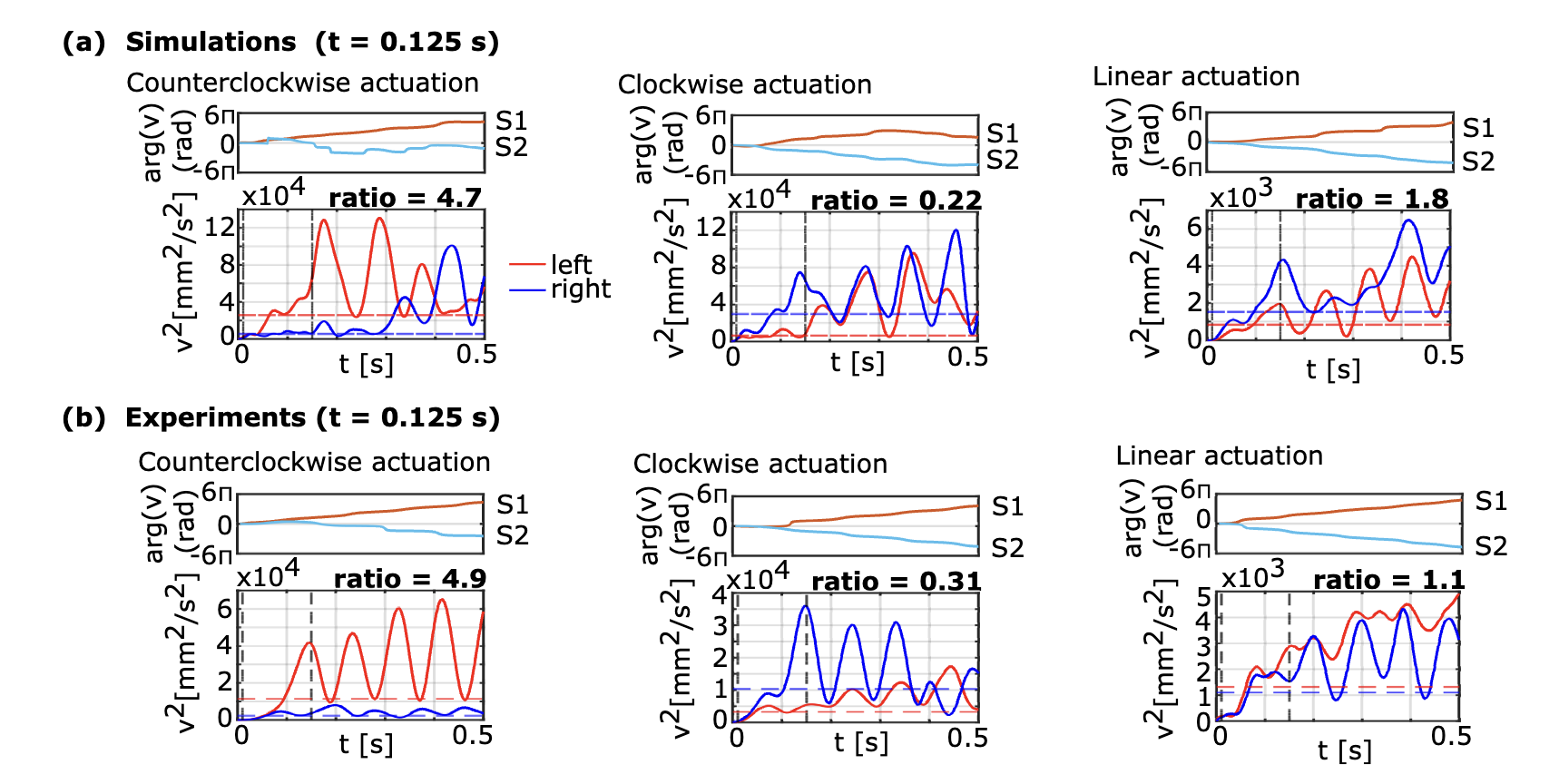}
    \caption{Experiments and simulation raw data for one-way edge waves in Maxwell lattices. {\bf (a)} Time domain simulations and {\bf (b)} experimental measurements of one-way edge waves in auxetic Maxwell lattices. In each panel, the top figures display the velocity vector phases ($arg(v)$, which are shifted to start from $0$) of the centroids of triangles $S1$ and $S2$ (as indicated in Fig.\ref{FigS_points_for_calculating_energy_ratio}) in the time domain under different excitation types. The bottom figures illustrate the average kinetic energy ($v^2$) in the same time range, calculated based on the triangles on the edge, as shown in Fig.~\ref{FigS_points_for_calculating_energy_ratio}. The term 'ratio' represents the energy ratio $E_{L}/E_{R}$.
    }
    \label{FigS_Energy ratio}
\end{figure*}

In Fig.~\ref{FigS_Energy ratio}, we present the raw data for the simulations and measurements of spin-momentum locking and one-way edge waves in Maxwell lattices, which were reported in the main text Fig.~4 (e-j). As explained in the main text, it was observed that the opposite propagating waves exhibit opposite spins. To further analyze this phenomenon, the velocity phases ($arg(v)$) of $S1$ and $S2$ were examined. It was consistently found that S1 rotates in a clockwise direction, while S2 rotates counterclockwise before the two edge waves encounter each other. This observation confirms the presence of spin-momentum locking in the edge waves. To quantify the energy distribution between the left- (L) and right- (R) propagating edge modes, the energy ratio ${E_L}/{E_R}$ was calculated. This ratio was obtained by integrating the average squared velocity (over the time range $0.01-0.15$ s) of the triangles located on the edge to the left and right of the excitation point, as depicted in Fig.~\ref{FigS_points_for_calculating_energy_ratio}. Considering that the group velocity at $5.5$ Hz is approximately $0.03$ m/s and the distance from the actuation source to the intersection of the two waves is around $0.45$ m, the two edge waves encounter each other after $0.15$ s.

\begin{widetext}
\section{Additional Maxwell lattice scattering simulations}

Additional FEM simulations are performed, where we deliberately introduce disorder in our simulated Maxwell lattice by removing or adding unit cells at the boundary or creating cavities within the bulk. 
As depicted in Fig.~\ref{FigS_defects}, the results demonstrate that this does not significantly disrupt the one-way edge transport along the edge.

\begin{figure*}[h]
    \centering
    \includegraphics[width=0.8\linewidth]{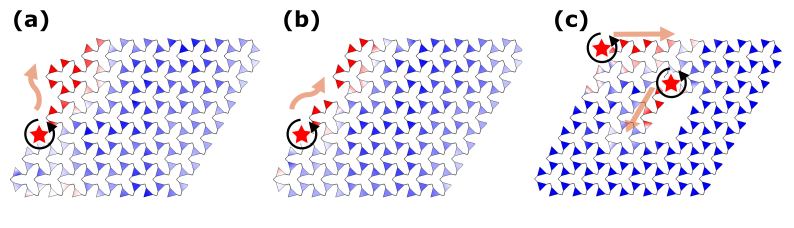}
    \caption{{\bf (a-c)} Robustness of edge wave transport in the auxetic Maxwell lattice by simulating various types of defects, where three triangles were added to the left edge ({\bf (a)}), three triangles were removed from the left edge ({\bf (b)}), and four triangles were removed from the bulk ({\bf (c)}). The excitation position is denoted by a red asterisk, and actuation is carried out in a counterclockwise direction.}
    \label{FigS_defects}
\end{figure*}
\end{widetext}


\end{document}